\documentclass[11pt]{article}

\usepackage{amsmath}
\usepackage{amssymb}
\usepackage{graphicx}
\usepackage{color}
\usepackage{cancel}
\usepackage{fancybox}
\usepackage{mathrsfs}
\usepackage{slashed}
\usepackage{capt-of}

\usepackage{axodraw4j}
\usepackage{pstricks}
\usepackage{color}

\usepackage{epsfig}

\newcommand{\ba}{\begin{array}{c}}
\newcommand{\ea}{\end{array}}
\newcommand{\bi}{\begin{itemize}}
\newcommand{\ei}{\end{itemize}}

\numberwithin{equation}{section}

\begin{document}
\begin{titlepage}

\begin{center}
\vspace*{15mm}
{\Large \bf Renormalization Group Evolution of the Non-Unitary operator
}
\vspace{.5in}

Subrata Khan\footnote{subrata@prl.res.in},

\vskip 0.5cm
{\small {\it Physical Research Laboratory, Navrangpura, 
Ahmedabad -380009, India}},\\

\vskip 1in

\end{center}

\begin{abstract}\noindent%
Integrating out a heavy field gives rise to effective Lagrangian containing
higher dimensional operators. In the context of Type-I seesaw mechanism, integrating out
the heavy right handed neutrino field leads to unique dimension five operator which gives
the tree level neutrino mass term. Apart from these there are dimension six operators that
can have important implications. A linear combination of two such operators gives rise to 
the non-unitarity in the lepton mixing matrix, $U_{\text{PMNS}}$. In this paper, we discuss the origin of 
non-unitarity at the high scale and its evolution through renormalization group running.
\end{abstract} 

\end{titlepage}


\newpage

\section{Introduction}
Standard Model (SM) of particle physics does not admit mass terms for the neutrinos. This was motivated by the absence of observational evidence of right handed neutrinos. However, in the last decades, neutrino interferometry experiments have observed oscillation of neutrinos. This is ascribed to neutrino mass and their mixing among different generations. Neutrino oscillation data gives neutrino mass squared differences, $\Delta m^2_{21}$ and $\left|\Delta m^2\right|_{31}$, to the order of $\mathcal{O}\left(10^{-5}\right)$ and $\left(10^{-3}\right)$ eV$^2$ respectively \cite{Schwetz:2008er}. Also from cosmology, we have an upper bound on the sum of neutrino masses \cite{hannestad} which is $\sum m_{i}\le 0.5$ eV. From the results of these observations it is clear that the neutrino masses are orders of magnitude smaller than other fermions {\it i.e.} charged leptons and quarks. A standard procedure to generate such a small neutrino mass naturally is to add heavy fields to the SM particle spectra. Below the mass scale of the heavy field, it gets integrated out from the theory and gives rise to the effective Lagrangian containing higher dimensional operators. These operators are {\it non-renormalizable}, and SM gauge group invariant. The lowest dimensional of such operators is the dimension-5 operator \cite{dimension5}.
\begin{eqnarray}
{\cal{L}}_5 \sim c^{(5)}_{ij} \left(\overline{l^{\mathcal{C}}_L}_i\epsilon \phi\right)\left(\phi^T \epsilon^T l_{L j}\right)\; 
\label{kappa-intro}  
\end{eqnarray}
where $l_L$ and $\phi$ are respectively the SM lepton and Higgs doublets.  As we can see from the above expression, $c^{(5)}$ has the dimension of $M^{-1}$, where $M$ is the mass of the right handed heavy particle. This operator violates lepton number by two units and leads to Majorana masses for neutrinos. After Spontaneous Symmetry Breaking (SSB), the $\phi$ field acquires Vacuum Expectation Value (VEV) $v$ and consequently neutrino mass becomes; $m_\nu \sim \frac{1}{2} c^{(5)} v^2$. With $v \sim 246$ GeV, to produce neutrino mass of $\sim 0.5$ eV, $M$ has to be of the order of ${\cal O}(10^{14}$ GeV). This way of generating small neutrino masses, by integrating out the heavy field is popularly called seesaw mechanism (Type-I) \cite{seesaw1}.

The next set of higher dimensional operators that arise in the theory are the dimension six operators. List of different dimension six operators can be found in \cite{dimension6}. A linear combination of two such operators is of special relevance in connection with the neutrino mass and mixings \cite{dimension6_1}. This operator is expressed as
\begin{eqnarray}
{\cal{L}}_6 \sim c^{(6)}_{ij} \left(\overline{l_{L}}_i\tilde{\phi}\right)i\slashed{\partial}\left(\tilde{\phi}^\dag l_{Lj}\right)\; 
\label{non-unin-intro}  
\end{eqnarray} 
Note that this operator is of Dirac type in contrast with the Majorana type operator (\ref{kappa-intro}). After symmetry breaking, the right hand side of Eq. (\ref{non-unin-intro}) becomes 
\begin{eqnarray}
c^{(6)}_{ij} \frac{v^2}{2}\overline{\nu_{L}}_i i\slashed{\partial} \nu_{Lj}\; 
\end{eqnarray}
which makes the coefficient of neutrino kinetic term deviate from unity by an amount of $\frac{1}{2}c^{(6)}v^2$. Now, in order to bring the kinetic term in its usual canonical form, the field $\nu_L$ is rescaled with a factor of $\left(1+\frac{1}{2}c^{(6)}v^2\right)^{1/2}$. This also modifies the charged current interaction. Consequently the lepton mixing matrix, $U_{\text{PMNS}}$, becomes non-unitary by a factor of $\left(1-\frac{1}{4}c^{(6)}v^2\right)$ \cite{hambye_non-unit}. Because of this, the operator in Eq. (\ref{non-unin-intro}) is called the non-unitary operator and $c^{(6)}$ is called the non-unitary parameter. The dimension of $c^{(6)}$ is $M^{-2}$. With $M$ is of the order of ${\cal O}(10^{14}$ GeV), non-unitarity is highly suppressed. 

Signature of such heavy neutrinos cannot be tested at colliders as their mass scale is way beyond the reach of these experiments. Therefore, from the point of view of testability at collider, one needs to bring down the seesaw scale to TeV. One way to circumvent this problem is provided by  inverse seesaw \cite{inv-seesaw}. In these scenarios ({\it i.e.} TeV scale seesaw \cite{inv-seesaw, adhikari, kers-smir}), mass scale of the heavy field is of the order of $\mathcal{O}(10^{3}$ GeV) and hence non-unitarity can be appreciable {\it i.e.} $c^{(6)}v^2\sim \mathcal{O}(10^{-2})$. Consequently the deviation of $U_{\text{PMNS}}$ from unitarity can be significant and can have testable consequences \cite{non-unin-test}.

As we know that these type of operators arise at comparatively higher scale whereas observations are made at a much lower scale \cite{antusch-non-unin}, the effect of Renormalization Group Evolution (RGE) needs to be incorporated. RGE for the dimension five operator, in the context of seesaw mechanism, is well known in the literature \cite{rg-study_seesaw, rg-eqn_gen}. RGE of the non-unitary operator is, however, not as widely studied. In \cite{dimension6_1} only part of the $\beta$-function of the non unitary parameter is computed. Here we shall present a detailed analysis of the origin of the non-unitary operator in the context of Type-I seesaw and one loop correction to that. Since the non-unitary operator becomes significant in TeV scale models, the RG running of this operator from the TeV scale to the electro-weak symmetry breaking scale is not expected to be as large as operators which are significant at very high energy ($10^{12}-10^{14}$ GeV). But as neutrino physics has entered the precision era, so strictly speaking, RGE of the non-unitary operator should be explored. Also for some cases, {\it e.g.}, where symmetry enforces $U_{e3}$ to be zero, a small RG running could evade the symmetry and may give a non-zero value of $U_{e3}$ as indicated by present data \cite{Schwetz:2008er}. 

The plan of the paper is as follows; in section 2 we shall present the Lagrangian, origin of non-unitarity at high scale and its renormalization procedure. In section 3 we give the one loop correction to the non-unitary operator, lepton doublet and Higgs doublet. Section 4 contains $\beta$-function of the non-unitary operator. We summarize in section 5. Appendix A contains the Feynman rules and appendix B contains the divergent part of different Feynman diagrams.
\section{Origin of non-unitarity and Renormalization procedure}
We add right handed gauge singlet heavy field to the SM particle spectra with mass is $M_j$.
The extended part of the Lagrangian is
\begin{eqnarray}
\label{lag}
{\cal{L}}_{\text{Yuk}} &=& -\left(Y_{\nu}\right)_{ij}\epsilon_{\alpha\beta}\bar{l}_{\alpha i}P_R N_{j}\phi_{\beta}^{\ast} - \left(Y_{\nu}^{\ast}\right)_{ij}\epsilon_{\alpha\beta}\phi_{\beta}\bar{N}_{j}P_L l_{\alpha i} \\
{\cal{L}}_{\text{Mass}} &=& -\frac{1}{2}M_{ij}\overline{N^{\mathcal{C}}_{i}}P_R N_j + \text{h.c.} \nonumber
\end{eqnarray}
where $\phi$ denotes $SU(2)$ higgs doublet with $\tilde{\phi} = \epsilon \phi^{*}$; $\epsilon$ is the $2\times 2$ anti-symmetric tensor. The lepton doublet of generation $i$ is denoted by $l_{i}$, and $N_{j}$ denote the heavy neutrino fields. The Yukawa coupling $Y_{\nu}$ is complex-valued $i\times j$ matrix in general. After SSB one gets the Dirac mass matrix for the neutrino as $m_D \sim v Y_\nu$. The Majorana mass matrix $M$ is $j \times j$ complex symmetric matrix.
\subsection{Origin of non-unitarity}
With the above Lagrangian (\ref{lag}) one can construct four possible dimension-6 Dirac type operators in the context of $l\phi\rightarrow l\phi$ and $l\phi^{\ast}\rightarrow l\phi^{\ast}$, two in each cases. Here we present one from each of them. The other two will be the same with fermion lines reversed.
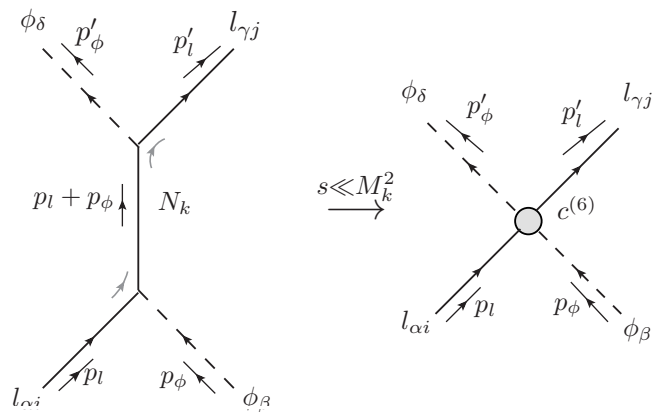
\begin{center}
\fcolorbox{white}{white}{
  \begin{picture}(263,160) (400,-251)
    \SetWidth{0.7}
    \SetColor{Black}
    \Line(443,-155)(443,-209)
    \Line[arrow,arrowpos=0.5,arrowlength=1.5,arrowwidth=0.6,arrowinset=0.2,flip](479,-118)(443,-154)
    \Line[arrow,arrowpos=0.5,arrowlength=1.5,arrowwidth=0.6,arrowinset=0.2](407,-246)(443,-210)
    \Line[dash,dashsize=5,arrow,arrowpos=0.5,arrowlength=1.5,arrowwidth=0.6,arrowinset=0.2,flip](407,-118)(443,-155)
    \Line[dash,dashsize=5,arrow,arrowpos=0.5,arrowlength=1.5,arrowwidth=0.6,arrowinset=0.2](479,-246)(443,-209)
    \Text(400,-112)[lb]{\small{\Black{$\phi_{\delta}$}}}
    \Text(479,-115)[lb]{\normalsize{\Black{$l_{\gamma j}$}}}
    \Text(397,-255)[lb]{\small{\Black{$l_{\alpha i}$}}}
    \Text(484,-256)[lb]{\small{\Black{$\phi_{\beta}$}}}
    \Text(452,-181)[lb]{\small{\Black{$N_{k}$}}}
    \SetWidth{0.6}
    \SetColor{Gray}
    \Arc[arrow,arrowpos=0.5,arrowlength=1.5,arrowwidth=0.6,arrowinset=0.2](419.5,-197.1)(20.106,-47.822,-14.105)
    \Arc[arrow,arrowpos=0.5,arrowlength=1.56,arrowwidth=0.624,arrowinset=0.2,clock](456,-160)(8.544,-159.444,-249.444)
    \Text(453,-248)[lb]{\small{\Black{$p_{\phi}$}}}
    \Text(458,-121)[lb]{\small{\Black{$p_l^{\prime}$}}}
    \Text(421,-120)[lb]{\normalsize{\Black{$p_{\phi}^{\prime}$}}}
    \SetWidth{0.5}
    \SetColor{Black}
    \Line[arrow,arrowpos=0.5,arrowlength=1.5,arrowwidth=0.6,arrowinset=0.2](426,-128)(414,-116)
    \Line[arrow,arrowpos=0.5,arrowlength=1.5,arrowwidth=0.6,arrowinset=0.2](460,-129)(474,-117)
    \Line[arrow,arrowpos=0.5,arrowlength=1.5,arrowwidth=0.6,arrowinset=0.2](472,-248)(460,-236)
    \Line[arrow,arrowpos=0.5,arrowlength=2,arrowwidth=0.8,arrowinset=0.2](413,-247)(426,-235)
    \Text(424,-246)[lb]{\small{\Black{$p_l$}}}
    \Text(405,-180)[lb]{\small{\Black{$p_l+p_{\phi}$}}}
    \SetWidth{0.7}
    \GOval(590,-183)(5,5)(0){0.882}
    \Line[arrow,arrowpos=0.5,arrowlength=1.5,arrowwidth=0.6,arrowinset=0.2,flip](624,-148)(593,-179)
    \Line[arrow,arrowpos=0.5,arrowlength=1.5,arrowwidth=0.6,arrowinset=0.2](555,-218)(587,-186)
    \Line[dash,dashsize=5,arrow,arrowpos=0.5,arrowlength=1.5,arrowwidth=0.6,arrowinset=0.2,flip](552,-146)(585,-179)
    \Line[dash,dashsize=5,arrow,arrowpos=0.5,arrowlength=1.5,arrowwidth=0.6,arrowinset=0.2](625,-218)(594,-187)
    \Text(512,-183)[lb]{\Large{\Black{$\overset{s \ll M_k^2}\longrightarrow$}}}
    \Text(544,-226)[lb]{\small{\Black{$l_{\alpha i}$}}}
    \Text(628,-230)[lb]{\small{\Black{$\phi_{\beta}$}}}
    \Text(544,-139)[lb]{\small{\Black{$\phi_{\delta}$}}}
    \Text(628,-141)[lb]{\small{\Black{$l_{\gamma j}$}}}
    \SetWidth{0.5}
    \Line[arrow,arrowpos=0.5,arrowlength=1.5,arrowwidth=0.6,arrowinset=0.2](603,-158)(619,-145)
    \Line[arrow,arrowpos=0.5,arrowlength=1.5,arrowwidth=0.6,arrowinset=0.2](573,-157)(559,-145)
    \Line[arrow,arrowpos=0.5,arrowlength=1.5,arrowwidth=0.6,arrowinset=0.2](559,-221)(572,-208)
    \Line[arrow,arrowpos=0.5,arrowlength=1.5,arrowwidth=0.6,arrowinset=0.2](620,-221)(606,-206)
    \Text(572,-220)[lb]{\small{\Black{$p_l$}}}
    \Text(602,-220)[lb]{\small{\Black{$p_{\phi}$}}}
    \Text(569,-147)[lb]{\small{\Black{$p_{\phi}^{\prime}$}}}
    \Text(605,-148)[lb]{\small{\Black{$p_l^{\prime}$}}}
    \Text(603,-183)[lb]{\small{\Black{$c^{(6)}$}}}
    \Line[arrow,arrowpos=0.5,arrowlength=1.5,arrowwidth=0.6,arrowinset=0.2](437,-185)(437,-169)
  \end{picture}
}
\captionof{figure}{\small {\it Feynman diagram, $l\phi\rightarrow l\phi$ mediated by heavy right handed neutrino (on the left hand side). On the right hand side, low momentum transfer approximation, leads to $c^{(6)}$ vertex.}}
\label{s-channel}
\end{center}
Tree level matching of both sides, in the limit $s\ll M_k^2$, gives
\begin{eqnarray}
\label{c6_1}
i\left(Y_{\nu}M^{-1}\left(M^{-1}\right)^{\ast}Y_{\nu}^{\dag}\right)_{ji}\epsilon_{\gamma\delta}\epsilon_{\alpha\beta}\left(\slashed{p}_l+\slashed{p}_{\phi} \right)P_L &=& \frac{i}{2}c^{(6)}_{ji}\epsilon_{\gamma\delta}\epsilon_{\alpha\beta}\left(\slashed{p}_l+\slashed{p}_{\phi} \right)P_L 
\end{eqnarray}
We identify $c^{(6)} = 2Y_{\nu}M^{-1}\left(M^{-1}\right)^{\ast}Y_{\nu}^{\dag}$. Note that $c^{(6)}$ is Hermitian. Here $s = \left(p_l + p_{\phi}\right)^2$. A second one is as follows
\begin{center}
\fcolorbox{white}{white}{
  \begin{picture}(318,104) (383,-109)
    \SetWidth{0.7}
    \SetColor{Black}
    \Line(425,-68)(478,-68)
    \Line[arrow,arrowpos=0.5,arrowlength=1.5,arrowwidth=0.6,arrowinset=0.2](390,-104)(425,-69)
    \Line[arrow,arrowpos=0.5,arrowlength=1.5,arrowwidth=0.6,arrowinset=0.2,flip](513,-33)(478,-68)
    \Line[dash,dashsize=5,arrow,arrowpos=0.5,arrowlength=1.5,arrowwidth=0.6,arrowinset=0.2](390,-33)(425,-68)
    \Line[dash,dashsize=5,arrow,arrowpos=0.5,arrowlength=1.5,arrowwidth=0.6,arrowinset=0.2,flip](513,-104)(478,-69)
    \GOval(628,-68)(6,6)(0){0.882}
    \Line[arrow,arrowpos=0.5,arrowlength=1.5,arrowwidth=0.6,arrowinset=0.2,flip](662,-33)(632,-64)
    \Line[arrow,arrowpos=0.5,arrowlength=1.5,arrowwidth=0.6,arrowinset=0.2](593,-103)(624,-72)
    \Line[dash,dashsize=5,arrow,arrowpos=0.5,arrowlength=1.5,arrowwidth=0.6,arrowinset=0.2](591,-32)(623,-64)
    \Line[dash,dashsize=5,arrow,arrowpos=0.5,arrowlength=1.5,arrowwidth=0.6,arrowinset=0.2,flip](664,-103)(633,-73)
    \Text(554,-71)[lb]{\Large{\Black{$\overset{t \ll M_k^2}\longrightarrow$}}}
    \SetWidth{0.6}
    \SetColor{Gray}
    \Arc[arrow,arrowpos=0.5,arrowlength=1.5,arrowwidth=0.6,arrowinset=0.2,clock](432.5,-87.5)(14.577,149.036,95.906)
    \Arc[arrow,arrowpos=0.5,arrowlength=1.5,arrowwidth=0.6,arrowinset=0.2](470.3,-53.3)(10.779,-96.927,-25.852)
    \Text(382,-26)[lb]{\small{\Black{$\phi_{\delta}$}}}
    \Text(516,-28)[lb]{\small{\Black{$l_{\gamma j}$}}}
    \Text(380,-114)[lb]{\small{\Black{$l_{\alpha i}$}}}
    \Text(516,-114)[lb]{\small{\Black{$\phi_{\beta}$}}}
    \Text(446,-59)[lb]{\small{\Black{$N_k$}}}
    \Text(584,-110)[lb]{\small{\Black{$l_{\alpha i}$}}}
    \Text(666,-114)[lb]{\small{\Black{$\phi_{\beta}$}}}
    \Text(584,-26)[lb]{\small{\Black{$\phi_{\delta}$}}}
    \Text(666,-28)[lb]{\small{\Black{$l_{\gamma j}$}}}
    \SetWidth{0.5}
    \SetColor{Black}
    \Line[arrow,arrowpos=0.5,arrowlength=1.5,arrowwidth=0.6,arrowinset=0.2](398,-105)(411,-91)
    \Line[arrow,arrowpos=0.5,arrowlength=1.5,arrowwidth=0.6,arrowinset=0.2](506,-104)(493,-91)
    \Line[arrow,arrowpos=0.5,arrowlength=1.5,arrowwidth=0.6,arrowinset=0.2](491,-44)(507,-30)
    \Line[arrow,arrowpos=0.5,arrowlength=1.5,arrowwidth=0.6,arrowinset=0.2](412,-44)(397,-30)
    \Line[arrow,arrowpos=0.5,arrowlength=1.5,arrowwidth=0.6,arrowinset=0.2](641,-44)(657,-30)
    \Line[arrow,arrowpos=0.5,arrowlength=1.5,arrowwidth=0.6,arrowinset=0.2](613,-43)(599,-31)
    \Line[arrow,arrowpos=0.5,arrowlength=1.5,arrowwidth=0.6,arrowinset=0.2](599,-105)(612,-91)
    \Line[arrow,arrowpos=0.5,arrowlength=2,arrowwidth=0.8,arrowinset=0.2](658,-105)(643,-92)
    \Text(408,-105)[lb]{\small{\Black{$p_l$}}}
    \Text(489,-104)[lb]{\small{\Black{$p_{\phi}$}}}
    \Text(407,-34)[lb]{\small{\Black{$p_{\phi}^{\prime}$}}}
    \Text(490,-35)[lb]{\small{\Black{$p_l^{\prime}$}}}
    \Text(441,-87)[lb]{\small{\Black{$p_l-p_{\phi}^{\prime}$}}}
    \Text(611,-104)[lb]{\small{\Black{$p_l$}}}
    \Text(640,-105)[lb]{\small{\Black{$p_{\phi}$}}}
    \Text(609,-34)[lb]{\small{\Black{$p_{\phi}^{\prime}$}}}
    \Text(643,-35)[lb]{\small{\Black{$p_l^{\prime}$}}}
    \Text(641,-68)[lb]{\small{\Black{$c^{(6)}$}}}
    \Line[arrow,arrowpos=0.5,arrowlength=1.5,arrowwidth=0.6,arrowinset=0.2](442,-74)(459,-74)
  \end{picture}
}
\captionof{figure}{\small {\it Feynman diagram, $l\phi^{\ast}\rightarrow l\phi^{\ast}$ mediated by heavy right handed neutrino (on the left hand side). On the right hand side, low momentum transfer approximation, leads to $c^{(6)}$ vertex.}}
\label{t-channel}
\end{center}
Tree level matching of both sides, in the limit $t\ll M_k^2$, gives
\begin{eqnarray}
\label{c6_2}
\left(Y_{\nu}M^{-1}\left(M^{-1}\right)^{\ast}Y_{\nu}^{\dag}\right)_{ji}\epsilon_{\gamma\beta}\epsilon_{\alpha\delta}\left(\slashed{p}_l-\slashed{p}_{\phi}^{\prime} \right)P_L &=& \frac{i}{2}c^{(6)}_{ji}\epsilon_{\gamma\beta}\epsilon_{\alpha\delta}\left(\slashed{p}_l-\slashed{p}_{\phi}^{\prime} \right)P_L 
\end{eqnarray}
Here $t = \left(p_l - p_{\phi}^{\prime}\right)^2$. Note that the external scalar lines as well as the momentum dependence in $s$-channel and $t$-channel are different. For the dimension-5 operator both $s$-channel and $t$-channel contribute to the same effective vertex. But here they contribute to different vertices. The effective dimension-6 operator (at low energy) in the Lagrangian is given by
\begin{eqnarray}
\label{l6}
{\cal{L}}^{(6)} = \frac{1}{2}c^{(6)}_{ji}\left(\overline{l_{L j}}\epsilon \phi^{\ast}\right)i\slashed{\partial}\left(\phi^T \epsilon^T l_{L i}\right)
\end{eqnarray}
\subsection{Renormalization procedure}
Here we consider the effect of incorporating one loop corrections to the non-unitary operator (\ref{l6}) {\it i.e. renormalization} of this operator. This requires wave function renormalization of $l_L$ and  $\phi$ doublet as well as the $c^{(6)}$ vertex renormalization. The wave function renormalization is defined as 
\begin{eqnarray}
\left(l_{L i}\right)_{B} &=& \left(Z_{l_L}^{\frac{1}{2}}\right)_{ij}l_{L j} \nonumber \\
\phi_B &=& Z_{\phi}^{\frac{1}{2}}\phi
\end{eqnarray}
The subscript $B$ stands for bare a quantity. Fields on the right hand side are the renormalized quantities. We shall follow the dimensional regularization procedure. $Z$'s are the renormalization constants which contain the divergence arising from  quantum corrections. Among different possibilities, we shall follow minimal subtraction (MS) scheme. In this scheme the divergent part of a loop diagram is absorbed in the counter term, defined as
\begin{eqnarray}
Z_i = 1 + \sum_{j\ge 1}\delta Z_{i,j}\frac{1}{\epsilon^j}
\end{eqnarray}
where $\epsilon$ is $4-d$ and $\delta Z_{,j}$ are coefficients of $\frac{1}{\epsilon}$ in the divergent part of a loop diagram. In the same spirit, the bare Lagrangian can be written as the sum of renormalized Lagrangian and the counter term
\begin{eqnarray}
\mathcal{L}^{(6)}_B = \mathcal{L}^{(6)} + \mathscr{L}^{(6)}
\end{eqnarray}
with $\mathscr{L}^{(6)}$, given by
\begin{eqnarray}
\mathscr{L}^{(6)} &=& \frac{1}{2}\Delta c^{(6)}_{ji}\left(\overline{l_{L j}}\epsilon \phi^{\ast}\right)i\slashed{\partial}\left(\phi^T \epsilon^T l_{L i}\right)
\end{eqnarray}
$\Delta c^{(6)}$ is the counterterm, related to the bare and renormalized coupling as below
\begin{eqnarray}
\label{cb-cr}
c^{(6)}_B &=& Z_{\phi}^{-\frac{1}{2}}Z_{l_L}^{-\frac{1}{2}}\left(c^{(6)}+\Delta c^{(6)}\right)\mu^{\epsilon}Z_{l_L}^{-\frac{1}{2}}Z_{\phi}^{-\frac{1}{2}}
\end{eqnarray}
where $\mu$ is the renormalization scale.
\section{One loop correction} 
In this section we shall present the one loop correction to the lepton and Higgs self energy as well as the $c^{(6)}$ vertex. Here we shall give only the one loop diagrams and the corresponding counter term. Expressions for individual diagram can be found in appendix B. In the MS scheme, the counterterm is determined by compensating the divergence arising from all possible one loop 1 particle irreducible diagrams. Pictorially it is shown below for $c^{(6)}$
\begin{center}
\fcolorbox{white}{white}{
  \begin{picture}(451,110) (2,-15)
    \SetWidth{0.5}
    \SetColor{Black}
    \Text(95.405,46.858)[lb]{\normalsize{\Black{$=$}}}
    \Text(211.496,42.215)[lb]{\normalsize{\Black{$+\sum $One-loop 1PI diagrams}}}
    \Text(357.559,42.637)[lb]{\normalsize{\Black{$+$}}}
    \Text(94.983,-17.73)[lb]{\normalsize{\Black{= UV finite}}}
    \SetWidth{0.7}
    \Line[arrow,arrowpos=0.5,arrowlength=1.5,arrowwidth=0.6,arrowinset=0.2,flip](73.032,79.364)(46.014,51.502)
    \Line[arrow,arrowpos=0.5,arrowlength=1.35,arrowwidth=0.54,arrowinset=0.2](9.287,15.197)(37.149,43.059)
    \Line[dash,dashsize=2.955,arrow,arrowpos=0.5,arrowlength=1.5,arrowwidth=0.6,arrowinset=0.2](9.287,78.942)(37.571,50.658)
    \Line[dash,dashsize=2.955,arrow,arrowpos=0.5,arrowlength=1.5,arrowwidth=0.6,arrowinset=0.2,flip](73.454,16.464)(45.592,43.481)
    \Text(1.689,8.865)[lb]{\normalsize{\Black{$l_{\alpha i}$}}}
    \Text(75.564,5.488)[lb]{\normalsize{\Black{$\phi_{\beta}$}}}
    \Text(1.689,85.696)[lb]{\normalsize{\Black{$\phi_{\delta}$}}}
    \Text(75.564,84.007)[lb]{\normalsize{\Black{$l_{\gamma j}$}}}
    \SetWidth{0.5}
    \Line[arrow,arrowpos=0.5,arrowlength=1.5,arrowwidth=0.6,arrowinset=0.2](53.191,69.232)(67.544,81.897)
    \Line[arrow,arrowpos=0.5,arrowlength=1.5,arrowwidth=0.6,arrowinset=0.2](27.862,70.076)(15.197,80.63)
    \Line[arrow,arrowpos=0.5,arrowlength=1.5,arrowwidth=0.6,arrowinset=0.2](14.775,13.509)(26.595,26.173)
    \Line[arrow,arrowpos=0.5,arrowlength=2,arrowwidth=0.8,arrowinset=0.2](69.654,14.775)(55.723,26.595)
    \Text(26.173,14.775)[lb]{\normalsize{\Black{$p_l$}}}
    \Text(52.346,13.087)[lb]{\normalsize{\Black{$p_{\phi}$}}}
    \Text(24.062,78.097)[lb]{\normalsize{\Black{$p_{\phi}^{\prime}$}}}
    \Text(54.457,78.942)[lb]{\normalsize{\Black{$p_l^{\prime}$}}}
    \SetWidth{0.7}
    \GBox(37.149,43.059)(45.592,51.502){0.4}
    \GOval(172.236,46.014)(5.066,5.066)(0){0.882}
    \Line[arrow,arrowpos=0.5,arrowlength=1.5,arrowwidth=0.6,arrowinset=0.2,flip](203.475,77.253)(176.458,49.813)
    \Line[arrow,arrowpos=0.5,arrowlength=1.5,arrowwidth=0.6,arrowinset=0.2](140.997,14.353)(168.437,41.793)
    \Line[dash,dashsize=2.955,arrow,arrowpos=0.5,arrowlength=1.5,arrowwidth=0.6,arrowinset=0.2](138.464,78.097)(166.326,50.236)
    \Line[dash,dashsize=2.955,arrow,arrowpos=0.5,arrowlength=1.5,arrowwidth=0.6,arrowinset=0.2,flip](203.475,14.775)(176.035,41.793)
    \Text(132.976,8.021)[lb]{\normalsize{\Black{$l_{\alpha i}$}}}
    \Text(206.43,5.066)[lb]{\normalsize{\Black{$\phi_{\beta}$}}}
    \Text(132.976,85.696)[lb]{\normalsize{\Black{$\phi_{\delta}$}}}
    \Text(206.43,84.007)[lb]{\normalsize{\Black{$l_{\gamma j}$}}}
    \SetWidth{0.5}
    \Line[arrow,arrowpos=0.5,arrowlength=1.5,arrowwidth=0.6,arrowinset=0.2](184.901,68.388)(198.409,80.63)
    \Line[arrow,arrowpos=0.5,arrowlength=1.5,arrowwidth=0.6,arrowinset=0.2](159.572,70.499)(147.329,81.474)
    \Line[arrow,arrowpos=0.5,arrowlength=1.5,arrowwidth=0.6,arrowinset=0.2](146.485,13.087)(158.305,25.329)
    \Line[arrow,arrowpos=0.5,arrowlength=1.5,arrowwidth=0.6,arrowinset=0.2](200.098,13.087)(187.433,24.907)
    \Text(157.039,13.931)[lb]{\normalsize{\Black{$p_l$}}}
    \Text(183.212,12.664)[lb]{\normalsize{\Black{$p_{\phi}$}}}
    \Text(155.772,76.831)[lb]{\normalsize{\Black{$p_{\phi}^{\prime}$}}}
    \Text(184.901,77.675)[lb]{\normalsize{\Black{$p_l^{\prime}$}}}
    \SetWidth{1.0}
    \Oval(405.262,45.592)(4.644,4.644)(0)
    \SetWidth{0.7}
    \Line[arrow,arrowpos=0.5,arrowlength=1.5,arrowwidth=0.6,arrowinset=0.2,flip](435.234,75.987)(408.639,48.969)
    \Line[arrow,arrowpos=0.5,arrowlength=1.5,arrowwidth=0.6,arrowinset=0.2](375.289,15.619)(402.306,42.637)
    \Line[dash,dashsize=2.955,arrow,arrowpos=0.5,arrowlength=1.5,arrowwidth=0.6,arrowinset=0.2](373.178,76.831)(401.884,48.125)
    \Line[dash,dashsize=2.955,arrow,arrowpos=0.5,arrowlength=1.5,arrowwidth=0.6,arrowinset=0.2,flip](436.5,14.353)(409.483,40.948)
    \Text(366.424,7.599)[lb]{\normalsize{\Black{$l_{\alpha i}$}}}
    \Text(439.455,4.221)[lb]{\normalsize{\Black{$\phi_{\beta}$}}}
    \Text(366.424,84.429)[lb]{\normalsize{\Black{$\phi_{\delta}$}}}
    \Text(439.455,80.208)[lb]{\normalsize{\Black{$l_{\gamma j}$}}}
    \SetWidth{0.5}
    \Line[arrow,arrowpos=0.5,arrowlength=1.5,arrowwidth=0.6,arrowinset=0.2](416.237,66.699)(430.59,79.786)
    \Line[arrow,arrowpos=0.5,arrowlength=1.5,arrowwidth=0.6,arrowinset=0.2](392.597,68.388)(379.511,80.63)
    \Line[arrow,arrowpos=0.5,arrowlength=1.5,arrowwidth=0.6,arrowinset=0.2](380.355,11.398)(393.019,24.485)
    \Line[arrow,arrowpos=0.5,arrowlength=1.5,arrowwidth=0.6,arrowinset=0.2](431.857,11.398)(417.926,24.062)
    \Text(390.486,13.087)[lb]{\normalsize{\Black{$p_l$}}}
    \Text(414.549,11.398)[lb]{\normalsize{\Black{$p_{\phi}$}}}
    \Text(387.953,75.564)[lb]{\normalsize{\Black{$p_{\phi}^{\prime}$}}}
    \Text(416.66,75.142)[lb]{\normalsize{\Black{$p_l^{\prime}$}}}
    \SetWidth{0.7}
    \Line(402.306,48.125)(408.217,42.215)
    \Line(402.306,42.637)(408.217,48.547)
  \end{picture}
}

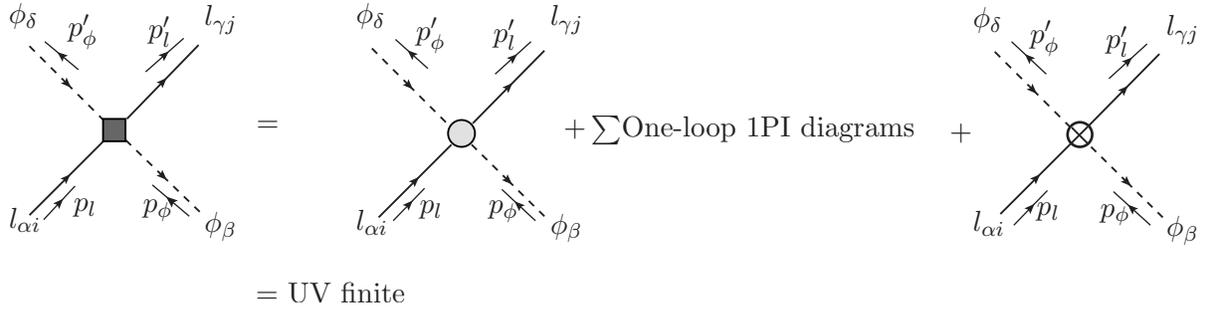
\captionof{figure}{\small {\it One loop renormalized $c^{(6)}$ vertex (on the left hand side). First diagram on the right hand side is the tree level $c^{(6)}$ vertex. Diagram on the extreme right is the counterterm vertex which eventually cancels the divergence.}}
\end{center}
the above figure can be expressed in terms of words as
\begin{eqnarray}
\label{compensate}
\text{Divergent part of all diagrams} \quad + \quad \text{the counterterm} &=& 0
\end{eqnarray}
Same procedure is followed to calculate the counterterms of $l_L$ and $\phi$ form the self energy diagrams.
\subsection{One loop correction of the lepton doublet $l_L$} 
The one loop self energy diagram of $l_L$ consists of the following diagrams \cite{kersten-thesis}
\begin{center}
\fcolorbox{white}{white}{
  \begin{picture}(451,72) (37,-62)
    \SetWidth{0.7}
    \SetColor{Black}
    \Line[arrow,arrowpos=0.5,arrowlength=1.5,arrowwidth=0.6,arrowinset=0.2](38.447,-15.379)(61.996,-15.379)
    \Line[arrow,arrowpos=0.5,arrowlength=1.5,arrowwidth=0.6,arrowinset=0.2](85.545,-15.379)(108.614,-15.379)
    \Arc[dash,dashsize=3.364,arrow,arrowpos=0.5,arrowlength=1.5,arrowwidth=0.6,arrowinset=0.2](73.771,-15.619)(11.777,1.169,178.831)
    \Arc(73.771,-15.139)(11.777,-178.831,-1.169)
    \Text(38.447,-11.534)[lb]{\small{\Black{$l_{\alpha i}$}}}
    \Text(100.924,-12.015)[lb]{\small{\Black{$l_{\gamma j}$}}}
    \Text(68.725,-37.967)[lb]{\small{\Black{$N_k$}}}
    \Text(70.647,-0.961)[lb]{\small{\Black{$\phi_{\beta}$}}}
    \SetWidth{0.5}
    \Arc[arrow,arrowpos=0.5,arrowlength=1.5,arrowwidth=0.6,arrowinset=0.2](75.677,-15.239)(7.173,-124.557,80.225)
    \Text(70.647,-12.495)[lb]{\small{\Black{$q$}}}
    \Text(65.36,-48.54)[lb]{\small{\Black{$p+q$}}}
    \Text(42.773,-26.913)[lb]{\small{\Black{$p$}}}
    \Text(101.405,-28.355)[lb]{\small{\Black{$p$}}}
    \SetWidth{0.6}
    \SetColor{Gray}
    \Arc[arrow,arrowpos=0.5,arrowlength=1.5,arrowwidth=0.6,arrowinset=0.2,clock](56.334,-24.092)(5.859,95.731,14.879)
    \Arc[arrow,arrowpos=0.5,arrowlength=1.5,arrowwidth=0.6,arrowinset=0.2,clock](90.832,-22.588)(4.806,-180,-270)
    \SetWidth{0.5}
    \SetColor{Black}
    \Line[arrow,arrowpos=0.5,arrowlength=1.5,arrowwidth=0.6,arrowinset=0.2](97.08,-18.743)(108.614,-18.743)
    \Line[arrow,arrowpos=0.5,arrowlength=1.5,arrowwidth=0.6,arrowinset=0.2](38.928,-18.262)(50.462,-18.262)
    \SetWidth{0.7}
    \Line[arrow,arrowpos=0.5,arrowlength=1.5,arrowwidth=0.6,arrowinset=0.2](160.998,-16.34)(184.547,-16.34)
    \Line[arrow,arrowpos=0.5,arrowlength=1.5,arrowwidth=0.6,arrowinset=0.2](208.096,-16.34)(231.165,-16.34)
    \Arc[dash,dashsize=3.364,arrow,arrowpos=0.5,arrowlength=1.5,arrowwidth=0.6,arrowinset=0.2](196.322,-16.58)(11.777,1.169,178.831)
    \Arc[arrow,arrowpos=0.5,arrowlength=1.5,arrowwidth=0.6,arrowinset=0.2](196.322,-16.1)(11.777,-178.831,-1.169)
    \Text(160.998,-12.495)[lb]{\small{\Black{$l_{\alpha i}$}}}
    \Text(223.475,-12.976)[lb]{\small{\Black{$l_{\gamma j}$}}}
    \Text(191.275,-38.928)[lb]{\small{\Black{$e_k$}}}
    \Text(193.198,-1.922)[lb]{\small{\Black{$\phi_{\beta}$}}}
    \SetWidth{0.5}
    \Arc[arrow,arrowpos=0.5,arrowlength=1.5,arrowwidth=0.6,arrowinset=0.2](198.228,-16.2)(7.173,-124.557,80.225)
    \Text(193.198,-13.457)[lb]{\small{\Black{$q$}}}
    \Text(187.911,-49.501)[lb]{\small{\Black{$p+q$}}}
    \Text(165.324,-27.874)[lb]{\small{\Black{$p$}}}
    \Text(223.956,-29.316)[lb]{\small{\Black{$p$}}}
    \SetWidth{0.6}
    \SetColor{Gray}
    \Arc[arrow,arrowpos=0.5,arrowlength=1.5,arrowwidth=0.6,arrowinset=0.2,clock](178.885,-25.053)(5.859,95.731,14.879)
    \Arc[arrow,arrowpos=0.5,arrowlength=1.5,arrowwidth=0.6,arrowinset=0.2,clock](213.383,-23.549)(4.806,-180,-270)
    \SetWidth{0.5}
    \SetColor{Black}
    \Line[arrow,arrowpos=0.5,arrowlength=1.5,arrowwidth=0.6,arrowinset=0.2](219.63,-19.704)(231.165,-19.704)
    \Line[arrow,arrowpos=0.5,arrowlength=1.5,arrowwidth=0.6,arrowinset=0.2](161.479,-19.224)(173.013,-19.224)
    \SetWidth{0.7}
    \Line[arrow,arrowpos=0.5,arrowlength=1.5,arrowwidth=0.6,arrowinset=0.2](285.471,-16.821)(309.02,-16.821)
    \Line[arrow,arrowpos=0.5,arrowlength=1.5,arrowwidth=0.6,arrowinset=0.2](332.569,-16.821)(355.638,-16.821)
    \Arc[arrow,arrowpos=0.5,arrowlength=1.5,arrowwidth=0.6,arrowinset=0.2](320.795,-16.58)(11.777,-178.831,-1.169)
    \Text(285.471,-12.976)[lb]{\small{\Black{$l_{\alpha i}$}}}
    \Text(347.948,-13.457)[lb]{\small{\Black{$l_{\gamma j}$}}}
    \Text(315.749,-39.409)[lb]{\small{\Black{$l_{\rho k}$}}}
    \Text(317.671,-0.961)[lb]{\small{\Black{$B$}}}
    \SetWidth{0.5}
    \Arc[arrow,arrowpos=0.5,arrowlength=1.5,arrowwidth=0.6,arrowinset=0.2](321.74,-16.681)(7.173,-124.557,80.225)
    \Text(317.671,-14.418)[lb]{\small{\Black{$q$}}}
    \Text(312.385,-49.982)[lb]{\small{\Black{$p+q$}}}
    \Text(289.797,-28.355)[lb]{\small{\Black{$p$}}}
    \Text(348.429,-29.797)[lb]{\small{\Black{$p$}}}
    \SetWidth{0.6}
    \SetColor{Gray}
    \Arc[arrow,arrowpos=0.5,arrowlength=1.5,arrowwidth=0.6,arrowinset=0.2,clock](303.358,-25.534)(5.859,95.731,14.879)
    \Arc[arrow,arrowpos=0.5,arrowlength=1.5,arrowwidth=0.6,arrowinset=0.2,clock](337.856,-24.03)(4.806,-180,-270)
    \SetWidth{0.5}
    \SetColor{Black}
    \Line[arrow,arrowpos=0.5,arrowlength=1.5,arrowwidth=0.6,arrowinset=0.2](344.104,-20.185)(355.638,-20.185)
    \Line[arrow,arrowpos=0.5,arrowlength=1.5,arrowwidth=0.6,arrowinset=0.2](285.952,-19.704)(297.486,-19.704)
    \SetWidth{0.7}
    \PhotonArc[clock](320.555,-16.821)(11.534,-180,-360){1.922}{6.5}
    \Line[arrow,arrowpos=0.5,arrowlength=1.5,arrowwidth=0.6,arrowinset=0.2](409.464,-17.782)(433.013,-17.782)
    \Line[arrow,arrowpos=0.5,arrowlength=1.5,arrowwidth=0.6,arrowinset=0.2](456.562,-17.782)(479.63,-17.782)
    \Arc[arrow,arrowpos=0.5,arrowlength=1.5,arrowwidth=0.6,arrowinset=0.2](444.788,-17.542)(11.777,-178.831,-1.169)
    \Text(409.464,-13.937)[lb]{\small{\Black{$l_{\alpha i}$}}}
    \Text(471.941,-14.418)[lb]{\small{\Black{$l_{\gamma j}$}}}
    \Text(439.741,-40.37)[lb]{\small{\Black{$l_{\rho k}$}}}
    \Text(441.664,-1.922)[lb]{\small{\Black{$W^a$}}}
    \SetWidth{0.5}
    \Arc[arrow,arrowpos=0.5,arrowlength=1.5,arrowwidth=0.6,arrowinset=0.2](445.732,-17.642)(7.173,-124.557,80.225)
    \Text(441.664,-15.379)[lb]{\small{\Black{$q$}}}
    \Text(436.377,-50.943)[lb]{\small{\Black{$p+q$}}}
    \Text(413.789,-29.316)[lb]{\small{\Black{$p$}}}
    \Text(472.422,-30.758)[lb]{\small{\Black{$p$}}}
    \SetWidth{0.6}
    \SetColor{Gray}
    \Arc[arrow,arrowpos=0.5,arrowlength=1.5,arrowwidth=0.6,arrowinset=0.2,clock](427.35,-26.495)(5.859,95.731,14.879)
    \Arc[arrow,arrowpos=0.5,arrowlength=1.5,arrowwidth=0.6,arrowinset=0.2,clock](461.849,-24.991)(4.806,-180,-270)
    \SetWidth{0.5}
    \SetColor{Black}
    \Line[arrow,arrowpos=0.5,arrowlength=1.5,arrowwidth=0.6,arrowinset=0.2](468.096,-21.146)(479.63,-21.146)
    \Line[arrow,arrowpos=0.5,arrowlength=1.5,arrowwidth=0.6,arrowinset=0.2](409.945,-20.665)(421.479,-20.665)
    \SetWidth{0.7}
    \PhotonArc[clock](444.547,-17.782)(11.534,-180,-360){1.922}{6.5}
    \Text(65.841,-63.438)[lb]{\footnotesize{\Black{(I)}}}
    \Text(188.873,-63.438)[lb]{\footnotesize{\Black{(II)}}}
    \Text(312.865,-64.88)[lb]{\footnotesize{\Black{(III)}}}
    \Text(436.858,-65.36)[lb]{\footnotesize{\Black{(IV)}}}
  \end{picture}
}

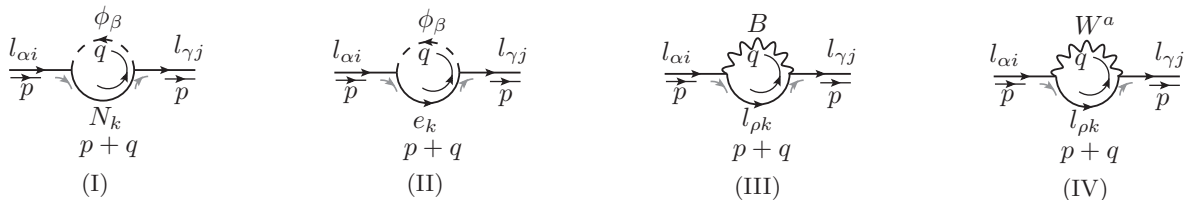
\captionof{figure}{\small {\it One loop self energy diagrams of $l_L$.}}
\label{lep-slf-enrg}
\end{center}
Taking the divergent part of the above diagrams and following Eq. (\ref{compensate}) we find the expression for the counterterm as
\begin{eqnarray}
\label{cntr-l_L}
\Delta Z_{l_L,1}&=&-\frac{1}{16\pi^2} \left[Y_{\nu}Y_{\nu}^{\dag} + Y_e Y_e^{\dag} + \frac{1}{2}\xi_B g_1^2 + \frac{3}{2}\xi_W g_2^2\right]
\end{eqnarray}
In the above expression, $Y_{e}$ is the Yukawa coupling coming from the $l-e-\phi$ vertex. $g_1$, $g_2$ are $U(1)_Y$ and $SU(2)_L$ gauge coupling constant. $\xi_B$, $\xi_W$ are the gauge fixing parameters for the $U(1)_Y$ and $SU(2)_L$ gauge fields respectively. $Y_{\nu}Y_{\nu}^{\dag}$ term contributes till the heavy field gets integrated out from the theory.
\subsection{One loop correction of the Higgs doublet $\phi$} 
The one loop self energy diagram of $\phi$ consists of the following diagrams \cite{kersten-thesis}
\begin{center}
\fcolorbox{white}{white}{
  \begin{picture}(451,86) (32,-31)
    \SetWidth{0.7}
    \SetColor{Black}
    \Line[dash,dashsize=3.665,arrow,arrowpos=0.5,arrowlength=1.5,arrowwidth=0.6,arrowinset=0.2](33.505,19.37)(58.634,19.37)
    \Line[dash,dashsize=3.665,arrow,arrowpos=0.5,arrowlength=1.5,arrowwidth=0.6,arrowinset=0.2](58.634,19.37)(83.763,19.37)
    \Arc[dash,dashsize=3.665,arrow,arrowpos=0.5,arrowlength=1.5,arrowwidth=0.6,arrowinset=0.2](59.158,29.841)(10.484,267,627)
    \SetWidth{0.5}
    \Arc[arrow,arrowpos=0.5,arrowlength=1.25,arrowwidth=0.5,arrowinset=0.2](60.251,30.455)(5.869,-85.329,74.94)
    \Line[arrow,arrowpos=0.5,arrowlength=1.5,arrowwidth=0.6,arrowinset=0.2](33.505,16.229)(44.499,16.229)
    \Line[arrow,arrowpos=0.5,arrowlength=1.5,arrowwidth=0.6,arrowinset=0.2](72.77,16.229)(84.287,16.229)
    \Text(30.888,23.558)[lb]{\small{\Black{$\phi_{\alpha}$}}}
    \Text(80.099,23.035)[lb]{\small{\Black{$\phi_{\beta}$}}}
    \Text(56.017,43.976)[lb]{\small{\Black{$\phi_{\gamma}$}}}
    \Text(54.97,30.364)[lb]{\small{\Black{$q$}}}
    \Text(37.17,7.329)[lb]{\small{\Black{$p$}}}
    \Text(76.958,7.329)[lb]{\small{\Black{$p$}}}
    \Line[dash,dashsize=3.665,arrow,arrowpos=0.5,arrowlength=1.25,arrowwidth=0.5,arrowinset=0.2](142.398,19.37)(168.05,19.37)
    \SetWidth{0.7}
    \Line[dash,dashsize=3.665,arrow,arrowpos=0.5,arrowlength=1.35,arrowwidth=0.54,arrowinset=0.2](192.656,19.37)(218.308,19.37)
    \Arc[dash,dashsize=3.665,arrow,arrowpos=0.5,arrowlength=1.35,arrowwidth=0.54,arrowinset=0.2](180.353,19.632)(12.306,-178.781,-1.219)
    \PhotonArc[clock](180.091,19.37)(12.565,-180,-360){2.094}{6.5}
    \Text(140.827,24.606)[lb]{\small{\Black{$\phi_{\alpha}$}}}
    \Text(209.932,24.082)[lb]{\small{\Black{$\phi_{\beta}$}}}
    \Text(175.38,-4.188)[lb]{\small{\Black{$\phi_{\gamma}$}}}
    \SetWidth{0.5}
    \Arc[arrow,arrowpos=0.5,arrowlength=1.25,arrowwidth=0.5,arrowinset=0.2](180.75,18.529)(5.627,-75.223,80.676)
    \Text(176.427,20.417)[lb]{\small{\Black{$q$}}}
    \Line[arrow,arrowpos=0.5,arrowlength=1.5,arrowwidth=0.6,arrowinset=0.2](207.838,15.706)(218.832,15.706)
    \Line[arrow,arrowpos=0.5,arrowlength=1.5,arrowwidth=0.6,arrowinset=0.2](142.921,16.229)(153.915,16.229)
    \Text(145.015,7.329)[lb]{\small{\Black{$p$}}}
    \Text(212.55,6.282)[lb]{\small{\Black{$p$}}}
    \Text(169.097,-12.565)[lb]{\small{\Black{$p+q$}}}
    \Text(175.38,38.741)[lb]{\small{\Black{$B$}}}
    \Line[dash,dashsize=3.665,arrow,arrowpos=0.5,arrowlength=1.25,arrowwidth=0.5,arrowinset=0.2](279.561,19.37)(305.213,19.37)
    \SetWidth{0.7}
    \Line[dash,dashsize=3.665,arrow,arrowpos=0.5,arrowlength=1.35,arrowwidth=0.54,arrowinset=0.2](329.819,19.37)(355.471,19.37)
    \Arc[dash,dashsize=3.665,arrow,arrowpos=0.5,arrowlength=1.35,arrowwidth=0.54,arrowinset=0.2](317.516,19.632)(12.306,-178.781,-1.219)
    \PhotonArc[clock](317.254,19.37)(12.565,-180,-360){2.094}{6.5}
    \Text(278.513,26.176)[lb]{\small{\Black{$\phi_{\alpha}$}}}
    \Text(347.095,25.129)[lb]{\small{\Black{$\phi_{\beta}$}}}
    \Text(310.972,-3.665)[lb]{\small{\Black{$\phi_{\gamma}$}}}
    \SetWidth{0.5}
    \Arc[arrow,arrowpos=0.5,arrowlength=1.25,arrowwidth=0.5,arrowinset=0.2](317.913,18.529)(5.627,-75.223,80.676)
    \Text(313.589,19.894)[lb]{\small{\Black{$q$}}}
    \Line[arrow,arrowpos=0.5,arrowlength=1.5,arrowwidth=0.6,arrowinset=0.2](345.001,15.706)(355.995,15.706)
    \Line[arrow,arrowpos=0.5,arrowlength=1.5,arrowwidth=0.6,arrowinset=0.2](280.084,16.229)(291.078,16.229)
    \Text(282.178,7.329)[lb]{\small{\Black{$p$}}}
    \Text(350.236,6.282)[lb]{\small{\Black{$p$}}}
    \Text(306.784,-12.565)[lb]{\small{\Black{$p+q$}}}
    \Text(312.542,38.217)[lb]{\small{\Black{$W^a$}}}
    \SetWidth{0.7}
    \Line[dash,dashsize=3.665,arrow,arrowpos=0.5,arrowlength=1.5,arrowwidth=0.6,arrowinset=0.2](418.817,18.847)(443.946,18.847)
    \Line[dash,dashsize=3.665,arrow,arrowpos=0.5,arrowlength=1.5,arrowwidth=0.6,arrowinset=0.2](443.946,18.847)(469.075,18.847)
    \SetWidth{0.5}
    \Arc[arrow,arrowpos=0.5,arrowlength=1.25,arrowwidth=0.5,arrowinset=0.2](443.403,30.344)(4.499,-54.105,144.105)
    \Line[arrow,arrowpos=0.5,arrowlength=1.5,arrowwidth=0.6,arrowinset=0.2](418.817,15.706)(429.811,15.706)
    \Line[arrow,arrowpos=0.5,arrowlength=1.5,arrowwidth=0.6,arrowinset=0.2](458.081,15.706)(469.599,15.706)
    \Text(416.723,22.511)[lb]{\small{\Black{$\phi_{\alpha}$}}}
    \Text(465.411,22.511)[lb]{\small{\Black{$\phi_{\beta}$}}}
    \Text(438.711,43.976)[lb]{\small{\Black{$B, W^a$}}}
    \Text(441.329,27.223)[lb]{\footnotesize{\Black{$q$}}}
    \Text(423.005,5.759)[lb]{\small{\Black{$p$}}}
    \Text(462.793,6.806)[lb]{\small{\Black{$p$}}}
    \SetWidth{0.7}
     \Bezier(444.47,19.894)(441.988,18.787)(441.664,19.059)(442.323,21.695) \Bezier(442.323,21.695)(443.226,24.2)(442.951,24.459)(440.505,23.407) \Bezier(440.505,23.407)(437.772,22.797)(437.447,23.16)(438.355,25.809) \Bezier(438.355,25.809)(439.729,28.125)(439.501,28.457)(436.846,28.009) \Bezier(436.846,28.009)(434.125,28.164)(433.957,28.56)(435.731,30.628) \Bezier(435.731,30.628)(438.161,32.028)(438.157,32.519)(435.701,33.873) \Bezier(435.701,33.873)(434.281,36.203)(434.511,36.568)(437.225,36.293) \Bezier(437.225,36.293)(439.601,34.877)(439.996,35.116)(439.838,37.878) \Bezier(439.838,37.878)(440.666,40.573)(441.155,40.685)(443.072,38.618) \Bezier(443.072,38.618)(444.245,36.16)(444.673,36.155)(445.906,38.583) \Bezier(445.906,38.583)(447.871,40.592)(448.349,40.466)(449.07,37.75) \Bezier(449.07,37.75)(448.726,35.018)(449.098,34.76)(451.533,36.047) \Bezier(451.533,36.047)(454.267,36.047)(454.463,35.656)(452.831,33.463) \Bezier(452.831,33.463)(450.329,32.387)(450.307,31.959)(452.685,30.631) \Bezier(452.685,30.631)(454.472,28.551)(454.303,28.141)(451.57,27.917) \Bezier(451.57,27.917)(448.887,28.282)(448.659,27.935)(450.061,25.619) \Bezier(450.061,25.619)(451.009,22.963)(450.684,22.581)(447.911,23.093) \Bezier(447.911,23.093)(445.416,24.056)(445.141,23.782)(446.093,21.283) \Bezier(446.093,21.283)(446.807,18.647)(446.482,18.358)(443.946,19.37)
    \Text(53.399,-32.982)[lb]{\footnotesize{\Black{(I)}}}
    \Text(174.333,-32.458)[lb]{\footnotesize{\Black{(II)}}}
    \Text(308.878,-32.982)[lb]{\footnotesize{\Black{(III)}}}
    \Text(437.664,-34.029)[lb]{\footnotesize{\Black{(IV)}}}
  \end{picture}
}
\end{center}
\begin{center}
\fcolorbox{white}{white}{
  \begin{picture}(451,77) (32,-132)
    \SetWidth{0.7}
    \SetColor{Black}
    \Line[dash,dashsize=3.677,arrow,arrowpos=0.5,arrowlength=1.35,arrowwidth=0.54,arrowinset=0.2](33.622,-83.531)(58.839,-83.531)
    \Line[dash,dashsize=3.677,arrow,arrowpos=0.5,arrowlength=1.35,arrowwidth=0.54,arrowinset=0.2](84.056,-83.005)(109.798,-83.005)
    \Arc[arrow,arrowpos=0.5,arrowlength=1.35,arrowwidth=0.54,arrowinset=0.2](71.71,-83.793)(12.349,1.219,178.781)
    \Arc(71.71,-82.743)(12.371,-176.348,-3.652)
    \SetWidth{0.5}
    \Line[arrow,arrowpos=0.5,arrowlength=1.5,arrowwidth=0.6,arrowinset=0.2](33.097,-86.683)(44.129,-86.683)
    \Line[arrow,arrowpos=0.5,arrowlength=1.5,arrowwidth=0.6,arrowinset=0.2](98.24,-86.157)(109.273,-86.157)
    \Arc[arrow,arrowpos=0.5,arrowlength=1.25,arrowwidth=0.5,arrowinset=0.2](72.399,-83.239)(8.195,-85.627,81.931)
    \SetWidth{0.6}
    \SetColor{Gray}
    \Arc[arrow,arrowpos=0.5,arrowlength=1.3,arrowwidth=0.52,arrowinset=0.2](69.609,-83.268)(7.097,141.009,218.991)
    \Text(32.046,-79.328)[lb]{\small{\Black{$\phi_{\alpha}$}}}
    \Text(104.545,-79.328)[lb]{\small{\Black{$\phi_{\beta}$}}}
    \Text(35.198,-96.139)[lb]{\normalsize{\Black{$p$}}}
    \Text(102.443,-95.088)[lb]{\small{\Black{$p$}}}
    \Text(68.295,-82.48)[lb]{\small{\Black{$q$}}}
    \Text(61.466,-114.526)[lb]{\small{\Black{$p+q$}}}
    \Text(64.093,-104.545)[lb]{\small{\Black{$N_k$}}}
    \Text(68.295,-66.719)[lb]{\small{\Black{$l_{\gamma m}$}}}
    \SetWidth{0.7}
    \SetColor{Black}
    \Line[dash,dashsize=3.677,arrow,arrowpos=0.5,arrowlength=1.35,arrowwidth=0.54,arrowinset=0.2](144.471,-83.531)(169.688,-83.531)
    \Line[dash,dashsize=3.677,arrow,arrowpos=0.5,arrowlength=1.35,arrowwidth=0.54,arrowinset=0.2](194.905,-83.005)(220.647,-83.005)
    \Arc[arrow,arrowpos=0.5,arrowlength=1.35,arrowwidth=0.54,arrowinset=0.2,flip](182.559,-83.793)(12.349,1.219,178.781)
    \Arc[arrow,arrowpos=0.5,arrowlength=1.35,arrowwidth=0.54,arrowinset=0.2,flip](182.559,-82.743)(12.371,-176.348,-3.652)
    \SetWidth{0.5}
    \Line[arrow,arrowpos=0.5,arrowlength=1.5,arrowwidth=0.6,arrowinset=0.2](143.946,-86.683)(154.978,-86.683)
    \Line[arrow,arrowpos=0.5,arrowlength=1.5,arrowwidth=0.6,arrowinset=0.2](209.089,-86.157)(220.122,-86.157)
    \Arc[arrow,arrowpos=0.5,arrowlength=1.25,arrowwidth=0.5,arrowinset=0.2](183.248,-83.239)(8.195,-85.627,81.931)
    \SetWidth{0.6}
    \SetColor{Gray}
    \Arc[arrow,arrowpos=0.5,arrowlength=1.3,arrowwidth=0.52,arrowinset=0.2](180.458,-83.268)(7.097,141.009,218.991)
    \Text(142.895,-79.328)[lb]{\small{\Black{$\phi_{\alpha}$}}}
    \Text(215.393,-79.328)[lb]{\small{\Black{$\phi_{\beta}$}}}
    \Text(146.047,-96.139)[lb]{\normalsize{\Black{$p$}}}
    \Text(213.292,-95.088)[lb]{\small{\Black{$p$}}}
    \Text(179.144,-82.48)[lb]{\small{\Black{$q$}}}
    \Text(172.315,-114.526)[lb]{\small{\Black{$p+q$}}}
    \Text(174.942,-104.545)[lb]{\small{\Black{$e_k$}}}
    \Text(179.144,-66.719)[lb]{\small{\Black{$l_{\gamma m}$}}}
    \SetWidth{0.7}
    \SetColor{Black}
    \Line[dash,dashsize=3.677,arrow,arrowpos=0.5,arrowlength=1.35,arrowwidth=0.54,arrowinset=0.2](271.606,-83.531)(296.823,-83.531)
    \Line[dash,dashsize=3.677,arrow,arrowpos=0.5,arrowlength=1.35,arrowwidth=0.54,arrowinset=0.2](322.04,-83.005)(347.782,-83.005)
    \Arc[arrow,arrowpos=0.5,arrowlength=1.35,arrowwidth=0.54,arrowinset=0.2](309.694,-83.793)(12.349,1.219,178.781)
    \Arc[arrow,arrowpos=0.5,arrowlength=1.35,arrowwidth=0.54,arrowinset=0.2](309.694,-82.743)(12.371,-176.348,-3.652)
    \SetWidth{0.5}
    \Line[arrow,arrowpos=0.5,arrowlength=1.5,arrowwidth=0.6,arrowinset=0.2](271.081,-86.683)(282.113,-86.683)
    \Line[arrow,arrowpos=0.5,arrowlength=1.5,arrowwidth=0.6,arrowinset=0.2](336.224,-86.157)(347.256,-86.157)
    \Arc[arrow,arrowpos=0.5,arrowlength=1.25,arrowwidth=0.5,arrowinset=0.2](310.382,-83.239)(8.195,-85.627,81.931)
    \SetWidth{0.6}
    \SetColor{Gray}
    \Arc[arrow,arrowpos=0.5,arrowlength=1.3,arrowwidth=0.52,arrowinset=0.2](307.592,-83.268)(7.097,141.009,218.991)
    \Text(270.03,-79.328)[lb]{\small{\Black{$\phi_{\alpha}$}}}
    \Text(342.528,-79.328)[lb]{\small{\Black{$\phi_{\beta}$}}}
    \Text(273.182,-96.139)[lb]{\normalsize{\Black{$p$}}}
    \Text(340.427,-95.088)[lb]{\small{\Black{$p$}}}
    \Text(306.279,-82.48)[lb]{\small{\Black{$q$}}}
    \Text(299.449,-114.526)[lb]{\small{\Black{$p+q$}}}
    \Text(302.076,-104.545)[lb]{\small{\Black{$u_k$}}}
    \Text(306.279,-66.719)[lb]{\small{\Black{$q_{\gamma m}$}}}
    \SetWidth{0.7}
    \SetColor{Black}
    \Line[dash,dashsize=3.677,arrow,arrowpos=0.5,arrowlength=1.35,arrowwidth=0.54,arrowinset=0.2](394.538,-83.531)(419.755,-83.531)
    \Line[dash,dashsize=3.677,arrow,arrowpos=0.5,arrowlength=1.35,arrowwidth=0.54,arrowinset=0.2](444.971,-83.005)(470.714,-83.005)
    \Arc[arrow,arrowpos=0.5,arrowlength=1.35,arrowwidth=0.54,arrowinset=0.2,flip](432.626,-83.793)(12.349,1.219,178.781)
    \Arc[arrow,arrowpos=0.5,arrowlength=1.35,arrowwidth=0.54,arrowinset=0.2,flip](432.626,-82.743)(12.371,-176.348,-3.652)
    \SetWidth{0.5}
    \Line[arrow,arrowpos=0.5,arrowlength=1.5,arrowwidth=0.6,arrowinset=0.2](394.012,-86.683)(405.045,-86.683)
    \Line[arrow,arrowpos=0.5,arrowlength=1.5,arrowwidth=0.6,arrowinset=0.2](459.156,-86.157)(470.188,-86.157)
    \Arc[arrow,arrowpos=0.5,arrowlength=1.25,arrowwidth=0.5,arrowinset=0.2](433.314,-83.239)(8.195,-85.627,81.931)
    \SetWidth{0.6}
    \SetColor{Gray}
    \Arc[arrow,arrowpos=0.5,arrowlength=1.3,arrowwidth=0.52,arrowinset=0.2](430.524,-83.268)(7.097,141.009,218.991)
    \Text(392.962,-79.328)[lb]{\small{\Black{$\phi_{\alpha}$}}}
    \Text(465.46,-79.328)[lb]{\small{\Black{$\phi_{\beta}$}}}
    \Text(396.114,-96.139)[lb]{\normalsize{\Black{$p$}}}
    \Text(463.359,-95.088)[lb]{\small{\Black{$p$}}}
    \Text(429.211,-82.48)[lb]{\small{\Black{$q$}}}
    \Text(422.381,-114.526)[lb]{\small{\Black{$p+q$}}}
    \Text(425.008,-104.545)[lb]{\small{\Black{$d_k$}}}
    \Text(429.211,-66.719)[lb]{\small{\Black{$q_{\gamma m}$}}}
    \Text(60.941,-135.54)[lb]{\footnotesize{\Black{(V)}}}
    \Text(171.264,-135.015)[lb]{\footnotesize{\Black{(VI)}}}
    \Text(301.026,-135.015)[lb]{\footnotesize{\Black{(VII)}}}
    \Text(423.957,-135.015)[lb]{\footnotesize{\Black{(VIII)}}}
  \end{picture}
}

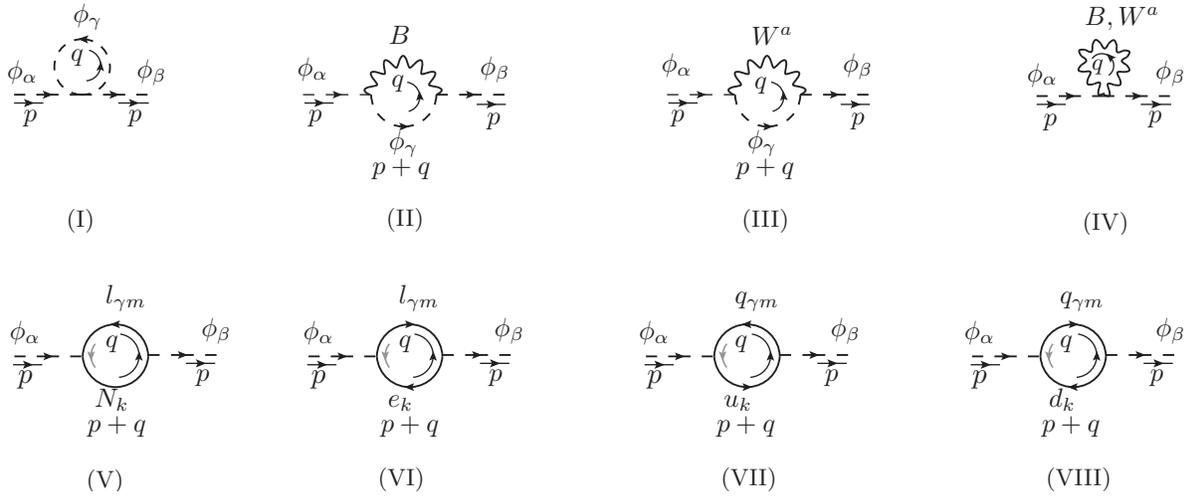
\captionof{figure}{\small {\it One loop self energy diagrams of $\phi$.}}
\label{higgs-slf-enrg}
\end{center}
Note that the fourth diagram (IV) vanishes in dimensional regularization, hence does not contribute. Following the similar procedure as above we find
\begin{eqnarray}
\label{cntr-higgs}
\Delta Z_{\phi,1}&=&-\frac{1}{16\pi^2} \left[2T - \frac{1}{2}\left(3-\xi_B\right)g_1^2 - \frac{3}{2}\left(3-\xi_W\right)g_2^2\right]
\end{eqnarray}
where, in general,
\begin{eqnarray}
T&=& \text{Tr}\left[Y_{\nu}Y_{\nu}^{\dag} + Y_e Y_e^{\dag} + 3Y_u Y_u^{\dag} + 3Y_d Y_d^{\dag}\right]
\end{eqnarray}
$Y_{u}$, $Y_{d}$ are the Yukawa coupling arising from the $q-u-\phi$, $q-d-\phi$ vertices respectively. Below the mass scale $M$, the heavy field $N$ gets decoupled from the theory. In that case, diagram V will not contribute to the self energy of $\phi$, consequently one needs to omit the term $Y_{\nu}Y_{\nu}^{\dag}$ from the expression of the $T$. Here we are not writing down the higgs mass counterterm because it is irrelevant for the present work.
\subsection{One loop correction of $c^{(6)}$ up to ${\cal{O}}(1/M^2)$}
Below we give all the one loop diagrams which contribute up to ${\cal{O}}(1/M^2)$ towards the correction of $c^{(6)}$. In the following, we proceed with the kind of vertex in Fig. \ref{t-channel}\footnote{Taking the other vertex, Fig. \ref{s-channel}, will give same $\beta$-function.}.
\begin{center}
\fcolorbox{white}{white}{
  \begin{picture}(451,125) (3,-23)
    \SetWidth{0.5}
    \SetColor{Black}
    \Text(55.936,-23.552)[lb]{\footnotesize{\Black{(I)}}}
    \Text(217.857,-24.814)[lb]{\footnotesize{\Black{(II)}}}
    \Text(381.46,-25.655)[lb]{\footnotesize{\Black{(III)}}}
    \SetWidth{0.7}
    \GOval(73.18,45.001)(3.365,3.785)(0){0.882}
    \Line[arrow,arrowpos=0.5,arrowlength=1.5,arrowwidth=0.6,arrowinset=0.2](75.703,47.945)(108.508,79.909)
    \Line[dash,dashsize=2.944,arrow,arrowpos=0.5,arrowlength=1.5,arrowwidth=0.6,arrowinset=0.2](76.124,42.057)(109.349,9.253)
    \Line[arrow,arrowpos=0.5,arrowlength=1.5,arrowwidth=0.6,arrowinset=0.2](103.882,7.991)(89.582,22.711)
    \Arc[arrow,arrowpos=0.5,arrowlength=1.5,arrowwidth=0.6,arrowinset=0.2,flip](54.371,43.091)(17.519,17.524,123.976)
    \Arc[dash,dashsize=2.944,arrow,arrowpos=0.5,arrowlength=1.5,arrowwidth=0.6,arrowinset=0.2,flip,clock](54.529,44.12)(17.215,-9.712,-118.696)
    \Arc[arrow,arrowpos=0.5,arrowlength=1.5,arrowwidth=0.6,arrowinset=0.2,clock](54.362,43.741)(16.978,-116.888,-234.823)
    \Line[dash,dashsize=2.944,arrow,arrowpos=0.5,arrowlength=1.5,arrowwidth=0.6,arrowinset=0.2](11.355,89.582)(44.16,56.777)
    \Line[arrow,arrowpos=0.5,arrowlength=1.5,arrowwidth=0.6,arrowinset=0.2](15.561,-5.047)(45.843,29.02)
    \SetWidth{0.5}
    \Arc[arrow,arrowpos=0.5,arrowlength=1.5,arrowwidth=0.6,arrowinset=0.2,clock](53.953,43.404)(9.885,-70.837,-284.073)
    \Text(60.563,34.908)[lb]{\small{\Black{$q$}}}
    \Line[arrow,arrowpos=0.5,arrowlength=1.5,arrowwidth=0.6,arrowinset=0.2](25.234,-4.626)(39.534,11.776)
    \Line[arrow,arrowpos=0.5,arrowlength=1.5,arrowwidth=0.6,arrowinset=0.2](34.487,76.965)(19.346,92.106)
    \Line[arrow,arrowpos=0.5,arrowlength=1.5,arrowwidth=0.6,arrowinset=0.2](86.218,70.236)(101.358,84.535)
    \SetWidth{0.6}
    \SetColor{Gray}
    \Arc[arrow,arrowpos=0.5,arrowlength=1.5,arrowwidth=0.6,arrowinset=0.2](31.336,26.772)(6.197,-95.874,44.592)
    \Arc[arrow,arrowpos=0.5,arrowlength=1.5,arrowwidth=0.6,arrowinset=0.2,clock](54.52,41.797)(14.499,149.265,102.79)
    \Text(5.467,-12.617)[lb]{\small{\Black{$l_{\alpha i}$}}}
    \Text(112.714,-0.841)[lb]{\small{\Black{$\phi_{\beta}$}}}
    \Text(110.19,84.115)[lb]{\small{\Black{$l_{\gamma j}$}}}
    \Text(2.523,92.947)[lb]{\small{\Black{$\phi_{\delta}$}}}
    \Text(24.814,36.59)[lb]{\small{\Black{$e_k$}}}
    \Text(66.451,58.88)[lb]{\small{\Black{$l_{\rho m}$}}}
    \Text(66.871,19.346)[lb]{\small{\Black{$\phi_{\eta}$}}}
    \Text(38.272,-1.262)[lb]{\small{\Black{$p_l$}}}
    \Text(82.853,3.785)[lb]{\small{\Black{$p_{\phi}$}}}
    \Text(87.9,85.376)[lb]{\small{\Black{$p_{l}^{\prime}$}}}
    \Text(27.337,90.423)[lb]{\small{\Black{$p_{\phi}^{\prime}$}}}
    \Text(6.309,47.104)[lb]{\small{\Black{$q+p_l$}}}
    \Text(37.431,68.974)[lb]{\small{\Black{$q+p_l-p_{\phi}^{\prime}$}}}
    \SetWidth{0.7}
    \SetColor{Black}
    \GOval(210.287,44.581)(3.365,3.365)(0){0.882}
    \Line[dash,dashsize=2.944,arrow,arrowpos=0.5,arrowlength=1.5,arrowwidth=0.6,arrowinset=0.2](174.959,79.909)(207.763,47.104)
    \Line[arrow,arrowpos=0.5,arrowlength=1.5,arrowwidth=0.6,arrowinset=0.2](174.538,8.832)(208.184,42.057)
    \Arc[dash,dashsize=2.944,arrow,arrowpos=0.5,arrowlength=1.5,arrowwidth=0.6,arrowinset=0.2,clock](231.638,43.812)(19.603,169.082,54.25)
    \Arc[arrow,arrowpos=0.5,arrowlength=1.5,arrowwidth=0.6,arrowinset=0.2](231.556,46.833)(19.245,-166.921,-54.721)
    \Arc[arrow,arrowpos=0.5,arrowlength=1.5,arrowwidth=0.6,arrowinset=0.2](230.197,45.561)(19.133,-50.952,45.9)
    \Line[dash,dashsize=2.944,arrow,arrowpos=0.5,arrowlength=1.5,arrowwidth=0.6,arrowinset=0.2](242.25,30.702)(272.532,-2.523)
    \Line[arrow,arrowpos=0.5,arrowlength=1.5,arrowwidth=0.6,arrowinset=0.2](243.512,59.301)(272.952,89.162)
    \SetWidth{0.5}
    \Arc[arrow,arrowpos=0.5,arrowlength=1.5,arrowwidth=0.6,arrowinset=0.2](233.037,44.61)(9.867,140.47,347.523)
    \SetWidth{0.6}
    \SetColor{Gray}
    \Arc[arrow,arrowpos=0.5,arrowlength=1.5,arrowwidth=0.6,arrowinset=0.2](233.208,41.987)(11.443,-80.48,-25.78)
    \Arc[arrow,arrowpos=0.5,arrowlength=1.5,arrowwidth=0.6,arrowinset=0.2,clock](254.107,60.301)(4.812,-95.975,-237.24)
    \Text(230.474,50.469)[lb]{\small{\Black{$q$}}}
    \Text(162.762,-1.262)[lb]{\small{\Black{$l_{\alpha i}$}}}
    \Text(279.681,-11.776)[lb]{\small{\Black{$\phi_{\beta}$}}}
    \Text(272.532,92.947)[lb]{\small{\Black{$l_{\gamma j}$}}}
    \Text(167.809,86.218)[lb]{\small{\Black{$\phi_{\delta}$}}}
    \Text(256.129,33.646)[lb]{\small{\Black{$e_k$}}}
    \Text(204.819,23.552)[lb]{\small{\Black{$l_{\rho m}$}}}
    \Text(215.334,65.609)[lb]{\small{\Black{$\phi_{\eta}$}}}
    \SetWidth{0.5}
    \SetColor{Black}
    \Line[arrow,arrowpos=0.5,arrowlength=1.5,arrowwidth=0.6,arrowinset=0.2](181.267,6.309)(195.146,20.608)
    \Line[arrow,arrowpos=0.5,arrowlength=1.5,arrowwidth=0.6,arrowinset=0.2](265.382,-3.785)(251.924,11.355)
    \Line[arrow,arrowpos=0.5,arrowlength=1.5,arrowwidth=0.6,arrowinset=0.2](251.082,79.068)(264.961,92.106)
    \Line[arrow,arrowpos=0.5,arrowlength=1.5,arrowwidth=0.6,arrowinset=0.2](196.408,68.554)(182.108,82.853)
    \Text(189.258,2.944)[lb]{\small{\Black{$p_l$}}}
    \Text(248.98,-7.991)[lb]{\small{\Black{$p_{\phi}$}}}
    \Text(251.924,90.844)[lb]{\small{\Black{$p_l^{\prime}$}}}
    \Text(191.361,78.647)[lb]{\small{\Black{$p_{\phi}^{\prime}$}}}
    \Text(259.494,44.581)[lb]{\small{\Black{$q+p_l^{\prime}$}}}
    \Text(201.455,11.355)[lb]{\small{\Black{$q+p_l^{\prime}-p_{\phi}$}}}
    \SetWidth{0.7}
    \GOval(362.535,44.16)(3.365,3.785)(0){0.882}
    \GOval(402.068,44.581)(3.365,3.785)(0){0.882}
    \Arc[dash,dashsize=2.944,arrow,arrowpos=0.5,arrowlength=1.5,arrowwidth=0.6,arrowinset=0.2,clock](382.301,46.628)(18.073,-17.423,-162.577)
    \Arc[arrow,arrowpos=0.5,arrowlength=1.5,arrowwidth=0.6,arrowinset=0.2](382.091,43.529)(17.905,12.894,167.106)
    \Line[arrow,arrowpos=0.5,arrowlength=1.5,arrowwidth=0.6,arrowinset=0.2](331.412,13.879)(360.011,42.057)
    \Line[arrow,arrowpos=0.5,arrowlength=1.5,arrowwidth=0.6,arrowinset=0.2](405.012,47.104)(433.191,75.703)
    \Line[dash,dashsize=2.944,arrow,arrowpos=0.5,arrowlength=1.5,arrowwidth=0.6,arrowinset=0.2](330.15,76.544)(359.591,46.684)
    \Line[dash,dashsize=2.944,arrow,arrowpos=0.5,arrowlength=1.5,arrowwidth=0.6,arrowinset=0.2](405.012,42.057)(434.032,11.355)
    \SetWidth{0.5}
    \Line[arrow,arrowpos=0.5,arrowlength=1.5,arrowwidth=0.6,arrowinset=0.2](338.562,11.355)(352.441,26.076)
    \Line[arrow,arrowpos=0.5,arrowlength=1.5,arrowwidth=0.6,arrowinset=0.2](428.565,9.673)(414.265,23.973)
    \Line[arrow,arrowpos=0.5,arrowlength=1.5,arrowwidth=0.6,arrowinset=0.2](412.583,64.768)(427.723,79.068)
    \Line[arrow,arrowpos=0.5,arrowlength=1.5,arrowwidth=0.6,arrowinset=0.2](353.702,62.665)(337.3,79.068)
    \SetWidth{0.6}
    \SetColor{Gray}
    \Arc[arrow,arrowpos=0.5,arrowlength=1.5,arrowwidth=0.6,arrowinset=0.2](339.333,61.814)(32.755,-62.33,-44.901)
    \Arc[arrow,arrowpos=0.5,arrowlength=1.5,arrowwidth=0.6,arrowinset=0.2](404.171,59.782)(5.107,-154.942,-25.058)
    \SetWidth{0.5}
    \SetColor{Black}
    \Arc[arrow,arrowpos=0.5,arrowlength=1.5,arrowwidth=0.6,arrowinset=0.2,clock](382.384,44.966)(7.67,146.432,-40.802)
    \Text(380.619,35.749)[lb]{\small{\Black{$q$}}}
    \Text(320.057,5.888)[lb]{\small{\Black{$l_{\alpha i}$}}}
    \Text(440.341,3.365)[lb]{\small{\Black{$\phi_{\beta}$}}}
    \Text(436.555,79.068)[lb]{\small{\Black{$l_{\gamma j}$}}}
    \Text(321.318,82.012)[lb]{\small{\Black{$\phi_{\delta}$}}}
    \Text(377.675,66.451)[lb]{\small{\Black{$l_{\rho k}$}}}
    \Text(382.301,11.776)[lb]{\small{\Black{$\phi_{\eta}$}}}
    \Text(350.338,13.038)[lb]{\small{\Black{$p_l$}}}
    \Text(409.639,8.411)[lb]{\small{\Black{$p_{\phi}$}}}
    \Text(420.574,84.956)[lb]{\small{\Black{$p_l^{\prime}$}}}
    \Text(340.665,82.432)[lb]{\small{\Black{$p_{\phi}^{\prime}$}}}
    \Text(360.011,78.647)[lb]{\small{\Black{$q+p_l-p_{\phi}^{\prime}$}}}
  \end{picture}
}
\end{center}
\begin{center}
\fcolorbox{white}{white}{
  \begin{picture}(451,137) (6,-172)
    \SetWidth{0.7}
    \SetColor{Black}
    \GOval(61.019,-96.137)(3.951,3.951)(0){0.882}
    \Line[arrow,arrowpos=0.5,arrowlength=1.5,arrowwidth=0.6,arrowinset=0.2](64.092,-93.504)(105.795,-51.8)
    \Line[arrow,arrowpos=0.5,arrowlength=1.5,arrowwidth=0.6,arrowinset=0.2](36.436,-120.282)(58.385,-99.21)
    \Line[dash,dashsize=3.073,arrow,arrowpos=0.5,arrowlength=1.5,arrowwidth=0.6,arrowinset=0.2](36.436,-71.993)(58.385,-93.065)
    \Line[dash,dashsize=3.073,arrow,arrowpos=0.5,arrowlength=1.5,arrowwidth=0.6,arrowinset=0.2](64.092,-99.21)(105.795,-140.914)
    \PhotonArc(47.367,-96.241)(27.179,114.73,244.246){1.756}{8.5}
    \Line[arrow,arrowpos=0.5,arrowlength=1.5,arrowwidth=0.6,arrowinset=0.2](15.803,-141.792)(37.753,-118.965)
    \Line[dash,dashsize=3.073,arrow,arrowpos=0.5,arrowlength=1.5,arrowwidth=0.6,arrowinset=0.2](14.925,-50.483)(35.997,-71.554)
    \SetWidth{0.5}
    \Line[arrow,arrowpos=0.5,arrowlength=1.5,arrowwidth=0.6,arrowinset=0.2](19.754,-144.426)(31.168,-133.012)
    \Line[arrow,arrowpos=0.5,arrowlength=1.5,arrowwidth=0.6,arrowinset=0.2](101.405,-143.548)(88.675,-130.817)
    \Line[arrow,arrowpos=0.5,arrowlength=1.5,arrowwidth=0.6,arrowinset=0.2](87.797,-61.019)(102.283,-47.41)
    \Line[arrow,arrowpos=0.5,arrowlength=1.5,arrowwidth=0.6,arrowinset=0.2](33.363,-59.702)(19.754,-46.971)
    \Arc[arrow,arrowpos=0.5,arrowlength=1.5,arrowwidth=0.6,arrowinset=0.2](41.018,-98.14)(10.841,84.044,278.294)
    \SetWidth{0.6}
    \SetColor{Gray}
    \Line[arrow,arrowpos=0.5,arrowlength=1.5,arrowwidth=0.6,arrowinset=0.2](36.436,-126.866)(41.703,-120.721)
    \Text(5.707,-151.449)[lb]{\small{\Black{$l_{\alpha i}$}}}
    \Text(108.868,-150.572)[lb]{\small{\Black{$\phi_{\beta}$}}}
    \Text(109.307,-46.532)[lb]{\small{\Black{$l_{\gamma j}$}}}
    \Text(4.829,-43.898)[lb]{\small{\Black{$\phi_{\delta}$}}}
    \Text(55.312,-117.209)[lb]{\small{\Black{$l_{\rho k}$}}}
    \Text(54.873,-83.846)[lb]{\small{\Black{$\phi_{\eta}$}}}
    \Text(5.707,-97.015)[lb]{\small{\Black{$B$}}}
    \Text(34.68,-99.649)[lb]{\small{\Black{$q$}}}
    \Text(29.412,-147.499)[lb]{\small{\Black{$p_l$}}}
    \Text(80.334,-147.06)[lb]{\small{\Black{$p_{\phi}$}}}
    \Text(47.849,-127.744)[lb]{\small{\Black{$q+p_l$}}}
    \Text(86.041,-46.971)[lb]{\small{\Black{$p_l^{\prime}$}}}
    \Text(28.095,-47.849)[lb]{\small{\Black{$p_{\phi}^{\prime}$}}}
    \Text(45.215,-70.237)[lb]{\small{\Black{$q+p_{\phi}^{\prime}$}}}
    \SetWidth{0.7}
    \SetColor{Black}
    \GOval(221.687,-97.454)(3.951,3.951)(0){0.882}
    \Line[arrow,arrowpos=0.5,arrowlength=1.5,arrowwidth=0.6,arrowinset=0.2](225.199,-94.382)(246.27,-73.31)
    \Line[arrow,arrowpos=0.5,arrowlength=1.5,arrowwidth=0.6,arrowinset=0.2](246.709,-72.871)(267.78,-51.8)
    \Line[dash,dashsize=3.073,arrow,arrowpos=0.5,arrowlength=1.5,arrowwidth=0.6,arrowinset=0.2](224.321,-100.527)(246.27,-121.599)
    \Line[dash,dashsize=3.073,arrow,arrowpos=0.5,arrowlength=1.5,arrowwidth=0.6,arrowinset=0.2](247.148,-122.477)(266.902,-142.231)
    \PhotonArc[clock](235.147,-97.554)(26.493,66.216,-65.175){1.756}{8.5}
    \Line[arrow,arrowpos=0.5,arrowlength=1.5,arrowwidth=0.6,arrowinset=0.2](175.155,-143.987)(219.053,-100.088)
    \Line[dash,dashsize=3.073,arrow,arrowpos=0.5,arrowlength=1.5,arrowwidth=0.6,arrowinset=0.2](175.155,-50.483)(217.736,-94.382)
    \SetWidth{0.5}
    \Arc[arrow,arrowpos=0.5,arrowlength=1.5,arrowwidth=0.6,arrowinset=0.2,clock](242.884,-97.072)(11.371,104.045,-92.847)
    \Line[arrow,arrowpos=0.5,arrowlength=1.5,arrowwidth=0.6,arrowinset=0.2](182.617,-144.426)(196.226,-130.817)
    \SetWidth{0.7}
    \Line[arrow,arrowpos=0.5,arrowlength=1.5,arrowwidth=0.6,arrowinset=0.2](262.951,-146.621)(248.465,-132.134)
    \Line[arrow,arrowpos=0.5,arrowlength=1.5,arrowwidth=0.6,arrowinset=0.2](248.465,-63.214)(262.073,-48.727)
    \SetWidth{0.5}
    \Line[arrow,arrowpos=0.5,arrowlength=1.5,arrowwidth=0.6,arrowinset=0.2](196.226,-64.092)(181.739,-48.288)
    \SetWidth{0.6}
    \SetColor{Gray}
    \Line[arrow,arrowpos=0.5,arrowlength=1.5,arrowwidth=0.6,arrowinset=0.2](240.124,-72.432)(245.392,-67.165)
    \Text(165.058,-151.888)[lb]{\small{\Black{$l_{\alpha i}$}}}
    \Text(274.804,-154.083)[lb]{\small{\Black{$\phi_{\beta}$}}}
    \Text(165.497,-44.776)[lb]{\small{\Black{$\phi_{\delta}$}}}
    \Text(272.17,-47.849)[lb]{\small{\Black{$l_{\gamma j}$}}}
    \Text(273.926,-95.26)[lb]{\small{\Black{$B$}}}
    \Text(243.636,-98.332)[lb]{\small{\Black{$q$}}}
    \Text(194.031,-146.182)[lb]{\small{\Black{$p_l$}}}
    \Text(245.831,-151.888)[lb]{\small{\Black{$p_{\phi}$}}}
    \Text(248.465,-49.166)[lb]{\small{\Black{$p_l^{\prime}$}}}
    \Text(190.958,-50.044)[lb]{\small{\Black{$p_{\phi}^{\prime}$}}}
    \Text(221.687,-80.334)[lb]{\small{\Black{$l_{\rho k}$}}}
    \Text(220.809,-117.209)[lb]{\small{\Black{$\phi_{\eta}$}}}
    \Text(212.029,-68.92)[lb]{\small{\Black{$q+p_l^{\prime}$}}}
    \Text(208.956,-130.378)[lb]{\small{\Black{$q+p_{\phi}$}}}
    \SetWidth{0.7}
    \SetColor{Black}
    \GOval(390.257,-97.893)(3.512,3.512)(0){0.882}
    \Line[arrow,arrowpos=0.5,arrowlength=1.5,arrowwidth=0.6,arrowinset=0.2](392.891,-95.26)(436.35,-52.678)
    \Line[dash,dashsize=3.073,arrow,arrowpos=0.5,arrowlength=1.5,arrowwidth=0.6,arrowinset=0.2,flip](387.623,-95.26)(343.285,-51.8)
    \Line[dash,dashsize=3.073,arrow,arrowpos=0.5,arrowlength=1.5,arrowwidth=0.6,arrowinset=0.2](393.33,-100.966)(413.962,-122.477)
    \Line[dash,dashsize=3.073,arrow,arrowpos=0.5,arrowlength=1.5,arrowwidth=0.6,arrowinset=0.2](415.279,-123.793)(435.911,-145.304)
    \Line[arrow,arrowpos=0.5,arrowlength=1.5,arrowwidth=0.6,arrowinset=0.2](365.674,-122.916)(388.062,-100.088)
    \Line[arrow,arrowpos=0.5,arrowlength=1.5,arrowwidth=0.6,arrowinset=0.2](344.602,-143.548)(365.235,-123.355)
    \PhotonArc(390.306,-92.832)(38.543,-129.724,-51.308){1.756}{8.5}
    \SetWidth{0.5}
    \Line[arrow,arrowpos=0.5,arrowlength=1.5,arrowwidth=0.6,arrowinset=0.2](349.431,-146.182)(359.528,-136.524)
    \Line[arrow,arrowpos=0.5,arrowlength=1.5,arrowwidth=0.6,arrowinset=0.2](432.399,-148.377)(420.986,-136.963)
    \Line[arrow,arrowpos=0.5,arrowlength=1.5,arrowwidth=0.6,arrowinset=0.2](416.157,-63.653)(429.765,-50.044)
    \Line[arrow,arrowpos=0.5,arrowlength=1.5,arrowwidth=0.6,arrowinset=0.2](363.918,-64.97)(350.309,-50.922)
    \Arc[arrow,arrowpos=0.5,arrowlength=1.5,arrowwidth=0.6,arrowinset=0.2,clock](391.574,-115.014)(9.229,-2.726,-177.274)
    \SetWidth{0.6}
    \SetColor{Gray}
    \Line[arrow,arrowpos=0.5,arrowlength=1.5,arrowwidth=0.6,arrowinset=0.2](359.089,-122.038)(364.796,-117.209)
    \Text(332.75,-152.766)[lb]{\small{\Black{$l_{\alpha i}$}}}
    \Text(442.057,-154.961)[lb]{\small{\Black{$\phi_{\beta}$}}}
    \Text(333.189,-47.41)[lb]{\small{\Black{$\phi_{\delta}$}}}
    \Text(440.301,-49.605)[lb]{\small{\Black{$l_{\gamma j}$}}}
    \Text(367.43,-102.722)[lb]{\small{\Black{$l_{\rho k}$}}}
    \Text(402.548,-104.917)[lb]{\small{\Black{$\phi_{\eta}$}}}
    \Text(391.574,-145.304)[lb]{\small{\Black{$B$}}}
    \Text(390.696,-118.526)[lb]{\small{\Black{$q$}}}
    \Text(341.53,-110.185)[lb]{\small{\Black{$q+p_l$}}}
    \Text(416.596,-111.941)[lb]{\small{\Black{$q-p_{\phi}$}}}
    \Text(358.65,-150.572)[lb]{\small{\Black{$p_l$}}}
    \Text(419.23,-154.522)[lb]{\small{\Black{$p_{\phi}$}}}
    \Text(359.528,-53.556)[lb]{\small{\Black{$p_{\phi}^{\prime}$}}}
    \Text(417.913,-50.922)[lb]{\small{\Black{$p_l^{\prime}$}}}
    \Text(43.02,-172.082)[lb]{\footnotesize{\Black{(IV)}}}
    \Text(212.029,-173.399)[lb]{\footnotesize{\Black{(V)}}}
    \Text(382.794,-174.277)[lb]{\footnotesize{\Black{(VI)}}}
  \end{picture}
}
\end{center}
\begin{center}
\fcolorbox{white}{white}{
  \begin{picture}(451,135) (11,-62)
    \SetWidth{0.7}
    \SetColor{Black}
    \GOval(67.845,10.067)(3.502,3.502)(0){0.882}
    \Line[arrow,arrowpos=0.5,arrowlength=1.5,arrowwidth=0.6,arrowinset=0.2](70.471,12.256)(92.356,34.579)
    \Line[arrow,arrowpos=0.5,arrowlength=1.5,arrowwidth=0.6,arrowinset=0.2](92.794,35.017)(114.679,57.34)
    \Line[dash,dashsize=3.064,arrow,arrowpos=0.5,arrowlength=1.5,arrowwidth=0.6,arrowinset=0.2,flip](65.656,12.256)(43.333,34.579)
    \Line[dash,dashsize=3.064,arrow,arrowpos=0.5,arrowlength=1.5,arrowwidth=0.6,arrowinset=0.2,flip](42.895,35.017)(20.572,57.34)
    \Line[arrow,arrowpos=0.5,arrowlength=1.5,arrowwidth=0.6,arrowinset=0.2,flip](65.656,7.441)(21.885,-36.33)
    \Line[dash,dashsize=3.064,arrow,arrowpos=0.5,arrowlength=1.5,arrowwidth=0.6,arrowinset=0.2](70.471,7.879)(114.679,-36.33)
    \PhotonArc[clock](67.364,5.245)(38.537,129.416,49.57){1.751}{8.5}
    \Line[arrow,arrowpos=0.5,arrowlength=1.5,arrowwidth=0.6,arrowinset=0.2](27.576,-38.956)(41.582,-25.825)
    \SetWidth{0.5}
    \Line[arrow,arrowpos=0.5,arrowlength=1.5,arrowwidth=0.6,arrowinset=0.2](109.427,-38.518)(96.733,-26.7)
    \Line[arrow,arrowpos=0.5,arrowlength=1.5,arrowwidth=0.6,arrowinset=0.2](96.733,46.835)(108.989,59.09)
    \Line[arrow,arrowpos=0.5,arrowlength=1.5,arrowwidth=0.6,arrowinset=0.2](40.707,46.835)(26.7,60.404)
    \SetWidth{0.6}
    \SetColor{Gray}
    \Line[arrow,arrowpos=0.5,arrowlength=1.5,arrowwidth=0.6,arrowinset=0.2](91.918,27.576)(98.046,34.141)
    \SetWidth{0.5}
    \SetColor{Black}
    \Arc[arrow,arrowpos=0.5,arrowlength=1.5,arrowwidth=0.6,arrowinset=0.2](68.001,28.802)(8.63,-26.514,158.589)
    \Text(65.656,25.387)[lb]{\small{\Black{$q$}}}
    \Text(63.905,55.589)[lb]{\small{\Black{$B$}}}
    \Text(12.256,-45.084)[lb]{\small{\Black{$l_{\alpha i}$}}}
    \Text(119.932,-46.835)[lb]{\small{\Black{$\phi_{\beta}$}}}
    \Text(117.305,60.404)[lb]{\small{\Black{$l_{\gamma j}$}}}
    \Text(14.007,64.343)[lb]{\small{\Black{$\phi_{\delta}$}}}
    \Text(84.04,10.505)[lb]{\small{\Black{$l_{\rho k}$}}}
    \Text(44.208,11.38)[lb]{\small{\Black{$\phi_{\eta}$}}}
    \Text(38.08,-40.707)[lb]{\small{\Black{$p_l$}}}
    \Text(89.73,-42.02)[lb]{\small{\Black{$p_{\phi}$}}}
    \Text(97.609,58.215)[lb]{\small{\Black{$p_l^{\prime}$}}}
    \Text(35.017,59.09)[lb]{\small{\Black{$p_{\phi}^{\prime}$}}}
    \Text(103.299,14.007)[lb]{\small{\Black{$q+p_l^{\prime}$}}}
    \Text(10.067,19.259)[lb]{\small{\Black{$q-p_{\phi}^{\prime}$}}}
    \SetWidth{0.7}
    \GOval(228.921,9.63)(3.502,3.502)(0){0.882}
    \Line[arrow,arrowpos=0.5,arrowlength=1.5,arrowwidth=0.6,arrowinset=0.2](231.109,11.818)(274.442,55.151)
    \Line[arrow,arrowpos=0.5,arrowlength=1.5,arrowwidth=0.6,arrowinset=0.2](182.961,-35.892)(226.294,7.003)
    \Line[dash,dashsize=3.064,arrow,arrowpos=0.5,arrowlength=1.5,arrowwidth=0.6,arrowinset=0.2](182.086,56.464)(204.409,33.703)
    \Line[dash,dashsize=3.064,arrow,arrowpos=0.5,arrowlength=1.5,arrowwidth=0.6,arrowinset=0.2](204.847,33.266)(226.294,11.818)
    \Line[dash,dashsize=3.064,arrow,arrowpos=0.5,arrowlength=1.5,arrowwidth=0.6,arrowinset=0.2](231.109,7.003)(252.994,-14.882)
    \Line[dash,dashsize=3.064,arrow,arrowpos=0.5,arrowlength=1.5,arrowwidth=0.6,arrowinset=0.2](253.87,-15.757)(274.88,-36.767)
    \PhotonArc(224.923,16.059)(27.345,138.606,232.691){1.751}{8.5}
    \PhotonArc[clock](235.182,6.427)(28.39,-49.995,-142.279){1.751}{8.5}
    \SetWidth{0.5}
    \Arc[arrow,arrowpos=0.5,arrowlength=1.5,arrowwidth=0.6,arrowinset=0.2,clock](232.801,-9.221)(6.308,77.438,-164.165)
    \SetWidth{0.7}
    \Line[arrow,arrowpos=0.5,arrowlength=1.5,arrowwidth=0.6,arrowinset=0.2](187.776,-37.643)(198.719,-26.7)
    \SetWidth{0.5}
    \Line[arrow,arrowpos=0.5,arrowlength=1.5,arrowwidth=0.6,arrowinset=0.2](270.503,-38.518)(259.122,-28.013)
    \Line[arrow,arrowpos=0.5,arrowlength=1.5,arrowwidth=0.6,arrowinset=0.2](254.745,45.084)(268.752,57.777)
    \Line[arrow,arrowpos=0.5,arrowlength=1.5,arrowwidth=0.6,arrowinset=0.2](200.47,46.835)(188.214,59.09)
    \Arc[arrow,arrowpos=0.5,arrowlength=1.5,arrowwidth=0.6,arrowinset=0.2,clock](210.564,13.562)(6.1,-111.301,-359.938)
    \Text(173.77,-43.771)[lb]{\small{\Black{$l_{\alpha i}$}}}
    \Text(279.695,-44.646)[lb]{\small{\Black{$\phi_{\beta}$}}}
    \Text(277.068,58.653)[lb]{\small{\Black{$l_{\gamma j}$}}}
    \Text(173.77,61.717)[lb]{\small{\Black{$\phi_{\delta}$}}}
    \Text(222.355,22.323)[lb]{\small{\Black{$\phi_{\rho}$}}}
    \Text(244.678,1.313)[lb]{\small{\Black{$\phi_{\eta}$}}}
    \Text(186.025,-4.815)[lb]{\small{\Black{$B$}}}
    \Text(222.355,-9.192)[lb]{\small{\Black{$q$}}}
    \Text(197.843,-39.394)[lb]{\small{\Black{$p_l$}}}
    \Text(252.119,-40.707)[lb]{\small{\Black{$p_{\phi}$}}}
    \Text(255.183,56.902)[lb]{\small{\Black{$p_l^{\prime}$}}}
    \Text(195.655,57.34)[lb]{\small{\Black{$p_{\phi}^{\prime}$}}}
    \Text(214.039,33.266)[lb]{\small{\Black{$q-p_{\phi}^{\prime}$}}}
    \Text(262.624,-1.313)[lb]{\small{\Black{$q-p_{\phi}$}}}
    \SetWidth{0.6}
    \SetColor{Gray}
    \Line[arrow,arrowpos=0.5,arrowlength=1.5,arrowwidth=0.6,arrowinset=0.2](215.352,-7.003)(221.48,-2.189)
    \SetWidth{0.5}
    \Line[arrow,arrowpos=0.5,arrowlength=1.667,arrowwidth=0.667,arrowinset=0.2](417.572,33.266)(422.825,38.08)
    \SetWidth{0.6}
    \Line[arrow,arrowpos=0.5,arrowlength=1.5,arrowwidth=0.6,arrowinset=0.2](368.111,-16.195)(374.239,-11.38)
    \SetWidth{0.7}
    \SetColor{Black}
    \GOval(397.438,8.754)(3.502,3.502)(0){0.882}
    \Line[arrow,arrowpos=0.5,arrowlength=1.5,arrowwidth=0.6,arrowinset=0.2](350.603,-37.643)(372.926,-15.757)
    \Line[arrow,arrowpos=0.5,arrowlength=1.5,arrowwidth=0.6,arrowinset=0.2](372.926,-15.757)(394.811,6.566)
    \Line[arrow,arrowpos=0.5,arrowlength=1.5,arrowwidth=0.6,arrowinset=0.2](399.626,10.943)(421.949,32.828)
    \Line[arrow,arrowpos=0.5,arrowlength=1.5,arrowwidth=0.6,arrowinset=0.2](421.949,32.828)(442.959,54.713)
    \Line[dash,dashsize=3.064,arrow,arrowpos=0.5,arrowlength=1.5,arrowwidth=0.6,arrowinset=0.2](351.916,54.276)(394.811,10.943)
    \Line[dash,dashsize=3.064,arrow,arrowpos=0.5,arrowlength=1.5,arrowwidth=0.6,arrowinset=0.2](399.626,6.566)(443.397,-37.205)
    \PhotonArc(393.588,3.667)(28.061,-137.419,-39.038){1.751}{8.5}
    \PhotonArc[clock](392.878,12.46)(35.749,35.59,-39.061){1.751}{8.5}
    \SetWidth{0.5}
    \Arc[arrow,arrowpos=0.5,arrowlength=1.5,arrowwidth=0.6,arrowinset=0.2,clock](396.051,-8.377)(8.407,-50.692,-201.816)
    \Arc[arrow,arrowpos=0.5,arrowlength=1.5,arrowwidth=0.6,arrowinset=0.2,clock](415.285,9.859)(7.061,156.33,-40.722)
    \Line[arrow,arrowpos=0.5,arrowlength=1.5,arrowwidth=0.6,arrowinset=0.2](355.856,-40.269)(367.674,-29.326)
    \SetWidth{0.7}
    \Line[arrow,arrowpos=0.5,arrowlength=1.5,arrowwidth=0.6,arrowinset=0.2](439.02,-39.831)(427.202,-27.576)
    \SetWidth{0.5}
    \Line[arrow,arrowpos=0.5,arrowlength=1.5,arrowwidth=0.6,arrowinset=0.2](426.326,46.397)(437.707,57.34)
    \Line[arrow,arrowpos=0.5,arrowlength=1.5,arrowwidth=0.6,arrowinset=0.2](370.738,42.895)(357.169,56.464)
    \Text(339.66,-45.959)[lb]{\small{\Black{$l_{\alpha i}$}}}
    \Text(447.336,-45.959)[lb]{\small{\Black{$\phi_{\beta}$}}}
    \Text(445.585,58.215)[lb]{\small{\Black{$l_{\gamma j}$}}}
    \Text(343.162,58.215)[lb]{\small{\Black{$\phi_{\delta}$}}}
    \Text(365.923,-5.252)[lb]{\small{\Black{$l_{\rho k}$}}}
    \Text(399.626,26.262)[lb]{\small{\Black{$l_{\eta m}$}}}
    \Text(398.313,-11.818)[lb]{\small{\Black{$q$}}}
    \Text(436.394,6.128)[lb]{\small{\Black{$B$}}}
    \Text(356.293,7.879)[lb]{\small{\Black{$q+p_l$}}}
    \Text(392.623,39.831)[lb]{\small{\Black{$q+p_l^{\prime}$}}}
    \Text(367.674,-41.582)[lb]{\small{\Black{$p_l$}}}
    \Text(421.512,-43.333)[lb]{\small{\Black{$p_{\phi}$}}}
    \Text(424.138,55.589)[lb]{\small{\Black{$p_l^{\prime}$}}}
    \Text(365.485,53.838)[lb]{\small{\Black{$p_{\phi}^{\prime}$}}}
    \Text(55.589,-63.905)[lb]{\footnotesize{\Black{(VII)}}}
    \Text(217.103,-64.343)[lb]{\footnotesize{\Black{(VIII)}}}
    \Text(390.872,-63.905)[lb]{\footnotesize{\Black{(IX)}}}
  \end{picture}
}
\end{center}
\begin{center}
\fcolorbox{white}{white}{
  \begin{picture}(163,139) (187,-172)
    \SetWidth{1.0}
    \SetColor{Black}
    \GOval(233,-99)(3,3)(0){0.882}
    \SetWidth{0.7}
    \Line[arrow,arrowpos=0.5,arrowlength=1.5,arrowwidth=0.6,arrowinset=0.2](194,-138)(231,-101)
    \Line[arrow,arrowpos=0.5,arrowlength=1.5,arrowwidth=0.6,arrowinset=0.2](231,-98)(192,-60)
    \Arc[dash,dashsize=5,arrow,arrowpos=0.5,arrowlength=1.5,arrowwidth=0.6,arrowinset=0.2](253,-96.464)(18.563,-165.857,-14.143)
    \Arc[dash,dashsize=5,arrow,arrowpos=0.5,arrowlength=1.5,arrowwidth=0.6,arrowinset=0.2,flip,clock](253.014,-97.986)(18.014,-179.955,-363.228)
    \Line[dash,dashsize=5,arrow,arrowpos=0.5,arrowlength=1.5,arrowwidth=0.6,arrowinset=0.2,flip](271,-99)(312,-59)
    \Line[dash,dashsize=5,arrow,arrowpos=0.5,arrowlength=1.5,arrowwidth=0.6,arrowinset=0.2](271,-99)(311,-139)
    \SetWidth{0.5}
    \Line[arrow,arrowpos=0.5,arrowlength=1.5,arrowwidth=0.6,arrowinset=0.2](200,-141)(208,-133)
    \Line[arrow,arrowpos=0.5,arrowlength=1.5,arrowwidth=0.6,arrowinset=0.2](305,-141)(297,-133)
    \Line[arrow,arrowpos=0.5,arrowlength=1.5,arrowwidth=0.6,arrowinset=0.2](295,-67)(304,-59)
    \Line[arrow,arrowpos=0.5,arrowlength=1.5,arrowwidth=0.6,arrowinset=0.2](207,-66)(198,-58)
    \Arc[arrow,arrowpos=0.5,arrowlength=1.5,arrowwidth=0.6,arrowinset=0.2,flip](252,-99.8)(10.121,119.604,322.224)
    \Text(252,-106)[lb]{\small{\Black{$q$}}}
    \Text(185,-151)[lb]{\small{\Black{$l_{\alpha i}$}}}
    \Text(313,-149)[lb]{\small{\Black{$\phi_{\beta}$}}}
    \Text(315,-57)[lb]{\small{\Black{$\phi_{\delta}$}}}
    \Text(184,-54)[lb]{\small{\Black{$l_{\gamma j}$}}}
    \Text(251,-131)[lb]{\small{\Black{$\phi_{\rho}$}}}
    \Text(264,-80)[lb]{\small{\Black{$\phi_{\eta}$}}}
    \Text(208,-145)[lb]{\small{\Black{$p_l$}}}
    \Text(289,-148)[lb]{\small{\Black{$p_{\phi}$}}}
    \Text(291,-59)[lb]{\small{\Black{$p_{\phi}^{\prime}$}}}
    \Text(208,-58)[lb]{\small{\Black{$p_l^{\prime}$}}}
    \Text(230,-68)[lb]{\small{\Black{$q+p_l-p_l^{\prime}$}}}
    \SetWidth{0.6}
    \SetColor{Gray}
    \Arc[arrow,arrowpos=0.5,arrowlength=1.5,arrowwidth=0.6,arrowinset=0.2](205.167,-100.5)(10.845,-43.755,43.755)
    \Text(245,-177)[lb]{\footnotesize{\Black{(X)}}}
  \end{picture}
}
\captionof{figure}{\small {\it Diagrams for one loop correction of $c^{(6)}$.}}
\label{c6-vrtx}
\end{center}
One loop correction of $c^{(5)}c^{\ast(5)}$ which is of the order of $\mathcal{O}\left(1/M^2\right)$ contributes towards the correction of $c^{(6)}$ (diagram III). There will be six additional gauge boson diagrams with $B$ replaced by $W^a$. Collecting the divergences from all diagrams and with the help of (\ref{compensate}), we find the expression for $\Delta c^{(6)}_{,1}$ as below
\begin{eqnarray}
\label{cntr-c6}
\Delta c^{(6)}_{,1} &=& -\frac{1}{16\pi^2} \left[2\left(c^{(6)}Y_e Y_e^{\dag} + Y_e Y_e^{\dag}c^{(6)}\right) + \frac{5}{2}c^{(5)}c^{\ast(5)}+ \left(\xi_B-\frac{3}{2}\right) g_1^2c^{(6)} \right . \nonumber \\
& & \left . + 3\left(\xi_W-\frac{1}{2}\right) g_2^2c^{(6)} \right]
\end{eqnarray}
In deriving Eq. (\ref{cntr-c6}), we have used the on-shell condition for the external fermions.
\section{$\beta$-function of $c^{(6)}$}
Now we are ready to compute the $\beta$-function of the parameter, $c^{(6)}$. We collect different expressions from the earlier section. From Eq. (\ref{cb-cr}), the bare non-unitary parameter $c^{(6)}_B$ is related to the renormalized one as
\begin{eqnarray}
c^{(6)}_B &=& Z_{\phi}^{-\frac{1}{2}}Z_{l_L}^{-\frac{1}{2}}\left(c^{(6)}+\Delta c^{(6)}\right)\mu^{\epsilon}Z_{l_L}^{-\frac{1}{2}}Z_{\phi}^{-\frac{1}{2}}\nonumber
\end{eqnarray}
To one loop order, $Z_{l_L}$, $Z_{\phi}$ and $c^{(6)}$ from (\ref{cntr-l_L}), (\ref{cntr-higgs}), (\ref{cntr-c6}) read
\begin{eqnarray}
\Delta Z_{l_L,1}&=&-\frac{1}{16\pi^2} \left[Y_e Y_e^{\dag} + \frac{1}{2}\xi_B g_1^2 + \frac{3}{2}\xi_B g_2^2\right]\\
\Delta Z_{\phi,1}&=&-\frac{1}{16\pi^2} \left[2T - \frac{1}{2}\left(3-\xi_B\right)g_1^2 - \frac{3}{2}\left(3-\xi_B\right)g_2^2\right]\\
\Delta c^{(6)}_{,1} &=& -\frac{1}{16\pi^2} \left[2\left(c^{(6)}Y_e Y_e^{\dag} + Y_e Y_e^{\dag}c^{(6)}\right) + \frac{5}{2}c^{(5)}c^{\ast(5)}+ \left(\xi_B-\frac{3}{2}\right) g_1^2c^{(6)} \right . \nonumber \\
& & \left . + 3\left(\xi_W-\frac{1}{2}\right) g_2^2c^{(6)} \right]
\end{eqnarray}
As we discussed earlier, $c^{(6)}$ gets generated below the mass scale $M$, when the heavy particle gets integrated out effectively implying that the term $Y_{\nu}Y_{\nu}^{\dag}$ can be dropped from the above expressions. Also the expression for $T$ becomes
\begin{eqnarray}
T&=& \text{Tr}\left[Y_e Y_e^{\dag} + 3Y_u Y_u^{\dag} + 3Y_d Y_d^{\dag}\right]
\end{eqnarray}
To compute the $\beta$-function we shall use the general equation for $\beta$-function, derived in \cite{rg-eqn_gen}. The $\beta$-function for the non-unitary parameter, $c^{(6)}$, works out to be
\begin{eqnarray}
\beta_{c^{(6)}}^{1-loop} &=& -\frac{1}{16\pi^2} \left[\frac{3}{2}\left(c^{(6)}Y_e Y_e^{\dag} + Y_e Y_e^{\dag}c^{(6)}\right) -2Tc^{(6)}+\frac{5}{2}c^{(5)}c^{\ast(5)}+ 3 g_2^2c^{(6)}\right]
\end{eqnarray}
This is the one loop $\beta$-function of the non-unitary parameter which governs the RGE of $c^{(6)}$. It is almost similar to the $\beta$-function of dimension-5 parameter, widely known as $\kappa$ in the literature \cite{rg-study_seesaw, rg-eqn_gen}. The term $c^{(5)}c^{\ast(5)}$ is additional here\footnote{Coefficient of this term differs from an earlier paper \cite{dimension6_1}.}.
\section{Summary}
In summary, we have discussed the origin of non-unitarity at high energy for type-I seesaw. Below the mass scale of the heavy particle, the dimension-6 Dirac type effective operator leads to the non-unitarity of the lepton mixing matrix. We have calculated the one loop correction and derived the $\beta$-function which governs RG evolution of the non-unitary operator. Since non-unitarity becomes significant in TeV scale models, running is low as compared to the operators originated at a higher scale. But in fine tuned situation this running may give significant modification to the result. RGE of non-unitary operator may also become relevant for precision measurement of neutrino mixings.\\

{\Large{\bf Acknowledgment}}\\

I sincerely thank S. Goswami for inspiring me to do this project and careful reading of the draft. I also thank N. Mahajan, S. Maji, S. Mohanty, S. Roy for useful discussions. I am also thankful to J. Kersten, A. Raychaudhuri for their feedback.
\section{Appendix A}
Relevant Feynman rules, those are used to compute above diagrams. Arrows on the particle line indicate charge flow. Arrows by the side of particle lines indicate momentum flow. Grey arrows indicate orientation, introduced in \cite{maj-feyn-rules}.\\
{\bf propagators:}
\begin{center}
\fcolorbox{white}{white}{
  \begin{picture}(388,35) (90,-83)
    \SetWidth{0.5}
    \SetColor{Black}
    \Text(225,-88)[lb]{\large{\Black{$\frac{i}{p^2-m^2+i\varepsilon}\delta_{\beta\alpha}$}}}
    \Text(87,-71)[lb]{\small{\Black{$\phi_{\alpha}$}}}
    \Text(200,-74)[lb]{\small{\Black{$\phi_{\beta}$}}}
    \SetWidth{0.7}
    \Line[dash,dashsize=6,arrow,arrowpos=0.5,arrowlength=1.5,arrowwidth=0.6,arrowinset=0.2](98,-77)(193,-77)
    \Text(443,-87)[lb]{\large{\Black{$i\frac{-g_{\mu\nu}+\left(1-\xi_{B,W}\right)\frac{p_{\mu}p_{\nu}}{p^2}}{p^2+i\varepsilon}\delta_{ab}\delta_{\beta\alpha}$}}}
    \Text(307,-69)[lb]{\small{\Black{$\left(B,W^a\right)_{\mu}$}}}
    \Text(395,-71)[lb]{\small{\Black{$\left(B,W^b\right)_{\nu}$}}}
    \Photon(321,-78)(418,-78){3.5}{7.5}
  \end{picture}
}
\end{center}
\begin{center}
\fcolorbox{white}{white}{
  \begin{picture}(451,45) (86,-81)
    \SetWidth{0.5}
    \SetColor{Black}
    \Text(248.682,-72.972)[lb]{\large{\Black{$\frac{i\left(\slashed{p}+M_j\right)}{p^2-M_j^2+i\varepsilon}\delta_{ji}$}}}
    \SetWidth{0.7}
    \Line(92.175,-62.41)(199.713,-62.41)
    \SetWidth{0.5}
    \Line[arrow,arrowpos=0.5,arrowlength=1.5,arrowwidth=0.6,arrowinset=0.2](126.741,-73.932)(162.267,-73.932)
    \SetWidth{0.6}
    \SetColor{Gray}
    \Line[arrow,arrowpos=0.5,arrowlength=1.5,arrowwidth=0.6,arrowinset=0.2](136.343,-50.889)(152.666,-50.889)
    \Text(83.534,-56.649)[lb]{\small{\Black{$N_{i}$}}}
    \Text(205.474,-58.57)[lb]{\small{\Black{$N_{j}$}}}
    \Text(140.183,-85.454)[lb]{\small{\Black{$p$}}}
    \Text(504.084,-73.932)[lb]{\large{\Black{$\frac{i\slashed{p}}{p^2+i\varepsilon}\delta_{ji}\delta_{\beta\alpha}$}}}
    \SetWidth{0.7}
    \SetColor{Black}
    \Line[arrow,arrowpos=0.5,arrowlength=1.5,arrowwidth=0.6,arrowinset=0.2](347.578,-63.371)(455.116,-63.371)
    \SetWidth{0.5}
    \Line[arrow,arrowpos=0.5,arrowlength=1.5,arrowwidth=0.6,arrowinset=0.2](382.144,-74.893)(417.67,-74.893)
    \SetWidth{0.6}
    \SetColor{Gray}
    \Line[arrow,arrowpos=0.5,arrowlength=1.5,arrowwidth=0.6,arrowinset=0.2](391.746,-51.849)(408.068,-51.849)
    \Text(338.937,-57.61)[lb]{\small{\Black{$f_{\alpha i}$}}}
    \Text(460.877,-59.53)[lb]{\small{\Black{$f_{\beta j}$}}}
    \Text(395.586,-86.414)[lb]{\small{\Black{$p$}}}
  \end{picture}
}
\end{center}
when momentum is opposite to the orientation, $p$ should be replaced by $-p$. $f$ denotes Dirac fermions. In case of singlet, the $SU(2)$ indices ({\it i.e.} $\alpha$, $\beta$) should be omitted.\\
{\bf $\phi^4$ interaction:}
\begin{center}
\fcolorbox{white}{white}{
  \begin{picture}(189,125) (105,-207)
    \SetWidth{0.5}
    \SetColor{Black}
    \Text(259,-156)[lb]{\small{\Black{$-\frac{i}{2}\mu^{\epsilon}\lambda\left(\delta_{\beta\alpha}\delta_{\gamma\delta}+\delta_{\beta\delta}\delta_{\gamma\alpha}\right)$}}}
    \SetWidth{0.7}
    \Line[dash,dashsize=7,arrow,arrowpos=0.5,arrowlength=1.5,arrowwidth=0.6,arrowinset=0.2](115,-111)(157,-153)
    \Line[dash,dashsize=7,arrow,arrowpos=0.5,arrowlength=1.5,arrowwidth=0.6,arrowinset=0.2](157,-153)(200,-196)
    \Line[dash,dashsize=7,arrow,arrowpos=0.5,arrowlength=1.5,arrowwidth=0.6,arrowinset=0.2](113,-196)(156,-153)
    \Line[dash,dashsize=7,arrow,arrowpos=0.5,arrowlength=1.5,arrowwidth=0.6,arrowinset=0.2](157,-152)(200,-109)
    \Text(102,-204)[lb]{\small{\Black{$\phi_{\alpha}$}}}
    \Text(207,-212)[lb]{\small{\Black{$\phi_{\beta}$}}}
    \Text(204,-103)[lb]{\small{\Black{$\phi_{\gamma}$}}}
    \Text(105,-106)[lb]{\small{\Black{$\phi_{\delta}$}}}
  \end{picture}
}
\end{center}
{\bf gauge boson interaction:}
\begin{center}
\fcolorbox{white}{white}{
  \begin{picture}(451,98) (86,-337)
    \SetWidth{0.5}
    \SetColor{Black}
    \Text(221.196,-302.923)[lb]{\small{\Black{$-\frac{i}{2}\mu^{\epsilon/2}g_1\left(p_{\mu}+p_{\mu}^{\prime}\right)\delta_{\beta\alpha}$}}}
    \Text(88.834,-341.122)[lb]{\small{\Black{$\phi_{\alpha}$}}}
    \Text(90.61,-257.618)[lb]{\small{\Black{$\phi_{\beta}$}}}
    \SetWidth{0.5}
    \Line[arrow,arrowpos=0.5,arrowlength=1.5,arrowwidth=0.6,arrowinset=0.2](92.387,-328.685)(108.377,-313.583)
    \Line[arrow,arrowpos=0.5,arrowlength=1.5,arrowwidth=0.6,arrowinset=0.2](109.266,-282.491)(95.052,-267.39)
    \Text(86.169,-317.137)[lb]{\small{\Black{$p$}}}
    \Text(83.504,-278.938)[lb]{\small{\Black{$p^{\prime}$}}}
    \Text(182.109,-283.38)[lb]{\small{\Black{$B_{\mu}$}}}
    \SetWidth{0.7}
    \Line[dash,dashsize=6.218,arrow,arrowpos=0.5,arrowlength=1.5,arrowwidth=0.6,arrowinset=0.2](101.271,-333.127)(136.804,-297.593)
    \Line[dash,dashsize=6.218,arrow,arrowpos=0.5,arrowlength=1.5,arrowwidth=0.6,arrowinset=0.2](135.916,-296.705)(100.382,-261.171)
    \Photon(135.916,-298.481)(213.201,-298.481){4.442}{8.5}
    \Text(506.353,-305.588)[lb]{\small{\Black{$-\frac{i}{2}\mu^{\epsilon/2}g_1\left(p_{\mu}+p_{\mu}^{\prime}\right)\sigma_{\beta\alpha}^a$}}}
    \Text(373.102,-342.01)[lb]{\small{\Black{$\phi_{\alpha}$}}}
    \Text(374.879,-258.506)[lb]{\small{\Black{$\phi_{\beta}$}}}
    \SetWidth{0.5}
    \Line[arrow,arrowpos=0.5,arrowlength=1.5,arrowwidth=0.6,arrowinset=0.2](376.655,-329.573)(392.645,-314.472)
    \Line[arrow,arrowpos=0.5,arrowlength=1.5,arrowwidth=0.6,arrowinset=0.2](393.534,-283.38)(379.32,-268.278)
    \Text(370.437,-318.025)[lb]{\small{\Black{$p$}}}
    \Text(367.772,-279.826)[lb]{\small{\Black{$p^{\prime}$}}}
    \Text(466.377,-284.268)[lb]{\small{\Black{$W_{\mu}^a$}}}
    \SetWidth{0.7}
    \Line[dash,dashsize=6.218,arrow,arrowpos=0.5,arrowlength=1.5,arrowwidth=0.6,arrowinset=0.2](385.539,-334.015)(421.072,-298.481)
    \Line[dash,dashsize=6.218,arrow,arrowpos=0.5,arrowlength=1.5,arrowwidth=0.6,arrowinset=0.2](420.184,-297.593)(384.65,-262.06)
    \Photon(421.072,-299.37)(498.357,-299.37){4.442}{8.5}
  \end{picture}
}
\end{center}
\begin{center}
\fcolorbox{white}{white}{
  \begin{picture}(451,113) (81,-32)
    \SetWidth{0.5}
    \SetColor{Black}
    \Text(215.515,12.408)[lb]{\small{\Black{$\frac{i}{2}\mu^{\epsilon/2}g_1\delta_{\beta\alpha}\delta_{ji}\gamma_{\mu}P_L$}}}
    \SetWidth{0.7}
    \Line[arrow,arrowpos=0.5,arrowlength=1.5,arrowwidth=0.6,arrowinset=0.2](92.084,-24.817)(134.534,17.633)
    \Line[arrow,arrowpos=0.5,arrowlength=1.5,arrowwidth=0.6,arrowinset=0.2](134.534,17.633)(91.431,60.083)
    \Photon(134.534,17.633)(203.107,17.633){3.265}{8}
    \Text(79.675,-35.266)[lb]{\small{\Black{$l_{\alpha i}$}}}
    \Text(79.675,67.267)[lb]{\small{\Black{$l_{\beta j}$}}}
    \Text(182.861,30.042)[lb]{\small{\Black{$B_{\mu}$}}}
    \SetWidth{0.6}
    \SetColor{Gray}
    \Arc[arrow,arrowpos=0.5,arrowlength=1.5,arrowwidth=0.6,arrowinset=0.2](103.263,16.783)(13.012,-69.793,60.266)
    \Text(510.052,11.102)[lb]{\small{\Black{$-\frac{i}{2}\mu^{\epsilon/2}g_1\delta_{\beta\alpha}\delta_{ji}\gamma_{\mu}P_L$}}}
    \SetWidth{0.7}
    \SetColor{Black}
    \Line[arrow,arrowpos=0.5,arrowlength=1.5,arrowwidth=0.6,arrowinset=0.2](390.539,-25.47)(432.989,16.98)
    \Line[arrow,arrowpos=0.5,arrowlength=1.5,arrowwidth=0.6,arrowinset=0.2](432.989,16.98)(389.886,59.43)
    \Photon(432.989,16.98)(501.562,16.98){3.265}{8}
    \Text(378.131,-35.919)[lb]{\small{\Black{$l_{\alpha i}$}}}
    \Text(378.131,66.614)[lb]{\small{\Black{$l_{\beta j}$}}}
    \Text(481.317,29.388)[lb]{\small{\Black{$B_{\mu}$}}}
    \SetWidth{0.6}
    \SetColor{Gray}
    \Arc[arrow,arrowpos=0.5,arrowlength=1.5,arrowwidth=0.6,arrowinset=0.2,flip](401.719,16.13)(13.012,-69.793,60.266)
  \end{picture}
}
\end{center}
\begin{center}
\fcolorbox{white}{white}{
  \begin{picture}(451,113) (81,-189)
    \SetWidth{0.5}
    \SetColor{Black}
    \Text(216.821,-144.983)[lb]{\small{\Black{$-\frac{i}{2}\mu^{\epsilon/2}g_2\sigma_{\beta\alpha}^a\delta_{ji}\gamma_{\mu}P_L$}}}
    \SetWidth{0.7}
    \Line[arrow,arrowpos=0.5,arrowlength=1.5,arrowwidth=0.6,arrowinset=0.2](92.084,-181.555)(134.534,-139.105)
    \Line[arrow,arrowpos=0.5,arrowlength=1.5,arrowwidth=0.6,arrowinset=0.2](134.534,-139.105)(91.431,-96.655)
    \Photon(134.534,-139.105)(203.107,-139.105){3.265}{8}
    \Text(79.675,-192.004)[lb]{\small{\Black{$l_{\alpha i}$}}}
    \Text(79.675,-89.471)[lb]{\small{\Black{$l_{\beta j}$}}}
    \Text(182.861,-126.697)[lb]{\small{\Black{$W_{\mu}^a$}}}
    \SetWidth{0.6}
    \SetColor{Gray}
    \Arc[arrow,arrowpos=0.5,arrowlength=1.5,arrowwidth=0.6,arrowinset=0.2](103.263,-139.955)(13.012,-69.793,60.266)
    \Text(510.052,-145.636)[lb]{\small{\Black{$\frac{i}{2}\mu^{\epsilon/2}g_2\sigma_{\beta\alpha}^a\delta_{ji}\gamma_{\mu}P_R$}}}
    \SetWidth{0.7}
    \SetColor{Black}
    \Line[arrow,arrowpos=0.5,arrowlength=1.5,arrowwidth=0.6,arrowinset=0.2](390.539,-182.208)(432.989,-139.758)
    \Line[arrow,arrowpos=0.5,arrowlength=1.5,arrowwidth=0.6,arrowinset=0.2](432.989,-139.758)(389.886,-97.308)
    \Photon(432.989,-139.758)(501.562,-139.758){3.265}{8}
    \Text(378.131,-192.657)[lb]{\small{\Black{$l_{\alpha i}$}}}
    \Text(378.131,-90.124)[lb]{\small{\Black{$l_{\beta j}$}}}
    \Text(481.317,-127.35)[lb]{\small{\Black{$W_{\mu}^a$}}}
    \SetWidth{0.6}
    \SetColor{Gray}
    \Arc[arrow,arrowpos=0.5,arrowlength=1.5,arrowwidth=0.6,arrowinset=0.2,flip](401.719,-140.608)(13.012,-69.793,60.266)
  \end{picture}
}
\end{center}
\begin{center}
\fcolorbox{white}{white}{
  \begin{picture}(451,71) (84,-76)
    \SetWidth{0.5}
    \SetColor{Black}
    \Text(176.288,-47.923)[lb]{\small{\Black{$\frac{i}{2}\mu^{\epsilon}g_2^2 g^{\mu\nu}\delta_{\alpha\beta}\delta_{ab}$}}}
    \SetWidth{0.7}
    \Line[dash,dashsize=3.423,arrow,arrowpos=0.5,arrowlength=1.5,arrowwidth=0.6,arrowinset=0.2](119.237,-46.782)(91.282,-19.968)
    \Line[dash,dashsize=3.423,arrow,arrowpos=0.5,arrowlength=1.5,arrowwidth=0.6,arrowinset=0.2,flip](119.237,-46.782)(91.853,-74.737)
    \Photon(119.237,-46.782)(147.192,-19.968){2.282}{5.5}
    \Photon(119.237,-46.782)(146.622,-74.737){2.282}{5.5}
    \Text(82.724,-77.59)[lb]{\small{\Black{$\phi_{\alpha}$}}}
    \Text(84.436,-17.115)[lb]{\small{\Black{$\phi_{\beta}$}}}
    \Text(149.474,-18.827)[lb]{\small{\Black{$W^b_{\nu}$}}}
    \Text(151.186,-79.301)[lb]{\small{\Black{$W^a_{\mu}$}}}
    \Text(349.724,-47.923)[lb]{\small{\Black{$\frac{i}{2}\mu^{\epsilon}g_1^2 g^{\mu\nu}\delta_{\alpha\beta}$}}}
    \Line[dash,dashsize=3.423,arrow,arrowpos=0.5,arrowlength=1.5,arrowwidth=0.6,arrowinset=0.2](292.673,-46.782)(264.718,-19.968)
    \Line[dash,dashsize=3.423,arrow,arrowpos=0.5,arrowlength=1.5,arrowwidth=0.6,arrowinset=0.2,flip](292.673,-46.782)(265.288,-74.737)
    \Photon(292.673,-46.782)(320.628,-19.968){2.282}{5.5}
    \Photon(292.673,-46.782)(320.058,-74.737){2.282}{5.5}
    \Text(256.16,-77.59)[lb]{\small{\Black{$\phi_{\alpha}$}}}
    \Text(257.872,-17.115)[lb]{\small{\Black{$\phi_{\beta}$}}}
    \Text(322.91,-18.827)[lb]{\small{\Black{$B_{\nu}$}}}
    \Text(324.622,-79.301)[lb]{\small{\Black{$B_{\mu}$}}}
    \Text(515.744,-48.494)[lb]{\small{\Black{$\frac{i}{2}\mu^{\epsilon}g_1g_2 g^{\mu\nu}\sigma_{\alpha\beta}^a$}}}
    \Line[dash,dashsize=3.423,arrow,arrowpos=0.5,arrowlength=1.5,arrowwidth=0.6,arrowinset=0.2](466.109,-46.782)(438.154,-19.968)
    \Line[dash,dashsize=3.423,arrow,arrowpos=0.5,arrowlength=1.5,arrowwidth=0.6,arrowinset=0.2,flip](466.109,-46.782)(438.724,-74.737)
    \Photon(466.109,-46.782)(494.064,-19.968){2.282}{5.5}
    \Photon(466.109,-46.782)(493.494,-74.737){2.282}{5.5}
    \Text(429.596,-77.59)[lb]{\small{\Black{$\phi_{\alpha}$}}}
    \Text(431.308,-17.115)[lb]{\small{\Black{$\phi_{\beta}$}}}
    \Text(496.346,-18.827)[lb]{\small{\Black{$B_{\nu}$}}}
    \Text(498.058,-79.301)[lb]{\small{\Black{$W^a_{\mu}$}}}
  \end{picture}
}
\end{center}
{\bf Yukawa interaction:}
\begin{center}
\fcolorbox{white}{white}{
  \begin{picture}(451,109) (72,-49)
    \SetWidth{0.5}
    \SetColor{Black}
    \Text(204.67,-6.259)[lb]{\small{\Black{$-i\mu^{\epsilon/2}\left(Y_{\nu}^{\ast}\right)_{ij}\epsilon_{\alpha\beta}P_L$}}}
    \SetWidth{0.7}
    \Line[arrow,arrowpos=0.5,arrowlength=1.5,arrowwidth=0.6,arrowinset=0.2](86.375,-44.439)(127.684,-3.13)
    \Line(127.058,-3.13)(85.749,38.18)
    \Line[dash,dashsize=4.381,arrow,arrowpos=0.5,arrowlength=1.5,arrowwidth=0.6,arrowinset=0.2](193.404,-3.13)(127.684,-3.13)
    \SetWidth{0.6}
    \SetColor{Gray}
    \Arc[arrow,arrowpos=0.5,arrowlength=1.5,arrowwidth=0.6,arrowinset=0.2](100.815,-3.18)(11.221,-70.691,63.755)
    \Text(70.727,-52.576)[lb]{\small{\Black{$l_{\alpha i}$}}}
    \Text(73.231,46.317)[lb]{\small{\Black{$N_j$}}}
    \Text(174.627,5.633)[lb]{\small{\Black{$\phi_{\beta}$}}}
    \Text(501.974,-5.633)[lb]{\small{\Black{$-i\mu^{\epsilon/2}\left(Y_{\nu}^{\ast}\right)_{ij}\epsilon_{\alpha\beta}P_L$}}}
    \SetWidth{0.7}
    \SetColor{Black}
    \Line[arrow,arrowpos=0.5,arrowlength=1.5,arrowwidth=0.6,arrowinset=0.2](386.182,-43.813)(427.491,-2.504)
    \Line(426.865,-2.504)(385.556,38.806)
    \Line[dash,dashsize=4.381,arrow,arrowpos=0.5,arrowlength=1.5,arrowwidth=0.6,arrowinset=0.2](492.585,-2.504)(426.865,-2.504)
    \SetWidth{0.6}
    \SetColor{Gray}
    \Arc[arrow,arrowpos=0.5,arrowlength=1.5,arrowwidth=0.6,arrowinset=0.2,flip](400.622,-2.554)(11.221,-70.691,63.755)
    \Text(370.534,-51.95)[lb]{\small{\Black{$l_{\alpha i}$}}}
    \Text(373.038,46.943)[lb]{\small{\Black{$N_j$}}}
    \Text(474.434,6.259)[lb]{\small{\Black{$\phi_{\beta}$}}}
  \end{picture}
}
\end{center}
\begin{center}
\fcolorbox{white}{white}{
  \begin{picture}(451,109) (72,-207)
    \SetWidth{0.5}
    \SetColor{Black}
    \Text(202.575,-162.684)[lb]{\small{\Black{$-i\mu^{\epsilon/2}\left(Y_{\nu}\right)_{ij}\epsilon_{\alpha\beta}P_R$}}}
    \SetWidth{0.7}
    \Line[arrow,arrowpos=0.5,arrowlength=1.5,arrowwidth=0.6,arrowinset=0.2,flip](86.017,-202.575)(127.155,-161.437)
    \Line(127.155,-161.437)(86.017,-120.299)
    \Line[dash,dashsize=4.363,arrow,arrowpos=0.5,arrowlength=1.5,arrowwidth=0.6,arrowinset=0.2,flip](192.602,-161.437)(127.155,-161.437)
    \SetWidth{0.6}
    \SetColor{Gray}
    \Arc[arrow,arrowpos=0.5,arrowlength=1.5,arrowwidth=0.6,arrowinset=0.2,flip](100.397,-161.487)(11.175,-70.691,63.755)
    \Text(70.434,-210.678)[lb]{\small{\Black{$l_{\alpha i}$}}}
    \Text(72.927,-112.196)[lb]{\small{\Black{$N_j$}}}
    \Text(173.903,-152.711)[lb]{\small{\Black{$\phi_{\beta}$}}}
    \Text(501.764,-161.437)[lb]{\small{\Black{$-i\mu^{\epsilon/2}\left(Y_{\nu}\right)_{ij}\epsilon_{\alpha\beta}P_R$}}}
    \SetWidth{0.7}
    \SetColor{Black}
    \Line[arrow,arrowpos=0.5,arrowlength=1.5,arrowwidth=0.6,arrowinset=0.2,flip](384.582,-201.952)(425.72,-160.814)
    \Line(425.72,-160.814)(384.582,-119.675)
    \Line[dash,dashsize=4.363,arrow,arrowpos=0.5,arrowlength=1.5,arrowwidth=0.6,arrowinset=0.2,flip](491.167,-160.814)(425.72,-160.814)
    \SetWidth{0.6}
    \SetColor{Gray}
    \Arc[arrow,arrowpos=0.5,arrowlength=1.5,arrowwidth=0.6,arrowinset=0.2](398.962,-160.864)(11.175,-70.691,63.755)
    \Text(368.999,-210.055)[lb]{\small{\Black{$l_{\alpha i}$}}}
    \Text(371.492,-111.572)[lb]{\small{\Black{$N_j$}}}
    \Text(472.468,-152.087)[lb]{\small{\Black{$\phi_{\beta}$}}}
  \end{picture}
}
\end{center}
\begin{center}
\fcolorbox{white}{white}{
  \begin{picture}(451,109) (72,-49)
    \SetWidth{0.5}
    \SetColor{Black}
    \Text(204.104,-4.993)[lb]{\small{\Black{$-i\mu^{\epsilon/2}\left(Y_{e}^{\ast}\right)_{ij}\delta_{\alpha\beta}P_L$}}}
    \SetWidth{0.7}
    \Line[arrow,arrowpos=0.5,arrowlength=1.5,arrowwidth=0.6,arrowinset=0.2](86.136,-44.316)(127.331,-3.121)
    \Line[arrow,arrowpos=0.5,arrowlength=1.5,arrowwidth=0.6,arrowinset=0.2](127.331,-3.745)(86.136,37.45)
    \Line[dash,dashsize=4.369,arrow,arrowpos=0.5,arrowlength=1.5,arrowwidth=0.6,arrowinset=0.2,flip](192.869,-3.121)(127.331,-3.121)
    \SetWidth{0.6}
    \SetColor{Gray}
    \Arc[arrow,arrowpos=0.5,arrowlength=1.5,arrowwidth=0.6,arrowinset=0.2](100.536,-3.171)(11.19,-70.691,63.755)
    \Text(70.531,-52.43)[lb]{\small{\Black{$l_{\alpha i}$}}}
    \Text(73.028,46.189)[lb]{\small{\Black{$e_j$}}}
    \Text(174.144,5.618)[lb]{\small{\Black{$\phi_{\beta}$}}}
    \Text(501.833,-3.121)[lb]{\small{\Black{$-i\mu^{\epsilon/2}\left(Y_{e}^{\ast}\right)_{ij}\delta_{\alpha\beta}P_L$}}}
    \SetWidth{0.7}
    \SetColor{Black}
    \Line[arrow,arrowpos=0.5,arrowlength=1.5,arrowwidth=0.6,arrowinset=0.2](385.113,-43.692)(426.309,-2.497)
    \Line[arrow,arrowpos=0.5,arrowlength=1.5,arrowwidth=0.6,arrowinset=0.2](426.309,-2.497)(385.113,38.699)
    \Line[dash,dashsize=4.369,arrow,arrowpos=0.5,arrowlength=1.5,arrowwidth=0.6,arrowinset=0.2,flip](491.847,-2.497)(426.309,-2.497)
    \SetWidth{0.6}
    \SetColor{Gray}
    \Arc[arrow,arrowpos=0.5,arrowlength=1.5,arrowwidth=0.6,arrowinset=0.2,flip](399.514,-2.547)(11.19,-70.691,63.755)
    \Text(369.509,-51.806)[lb]{\small{\Black{$l_{\alpha i}$}}}
    \Text(372.006,46.813)[lb]{\small{\Black{$e_j$}}}
    \Text(473.122,6.242)[lb]{\small{\Black{$\phi_{\beta}$}}}
  \end{picture}
}
\end{center}
\begin{center}
\fcolorbox{white}{white}{
  \begin{picture}(451,109) (72,-208)
    \SetWidth{0.5}
    \SetColor{Black}
    \Text(203.701,-163.587)[lb]{\small{\Black{$-i\mu^{\epsilon/2}\left(Y_{e}\right)_{ij}\delta_{\alpha\beta}P_R$}}}
    \SetWidth{0.7}
    \Line[arrow,arrowpos=0.5,arrowlength=1.5,arrowwidth=0.6,arrowinset=0.2,flip](86.494,-203.701)(127.861,-162.334)
    \Line[arrow,arrowpos=0.5,arrowlength=1.5,arrowwidth=0.6,arrowinset=0.2,flip](127.235,-162.334)(85.868,-120.967)
    \Line[dash,dashsize=4.387,arrow,arrowpos=0.5,arrowlength=1.5,arrowwidth=0.6,arrowinset=0.2](193.672,-162.334)(127.861,-162.334)
    \SetWidth{0.6}
    \SetColor{Gray}
    \Arc[arrow,arrowpos=0.5,arrowlength=1.5,arrowwidth=0.6,arrowinset=0.2,flip](100.955,-162.384)(11.237,-70.691,63.755)
    \Text(70.825,-211.849)[lb]{\small{\Black{$l_{\alpha i}$}}}
    \Text(73.332,-112.819)[lb]{\small{\Black{$e_j$}}}
    \Text(174.869,-153.559)[lb]{\small{\Black{$\phi_{\beta}$}}}
    \Text(502.044,-162.334)[lb]{\small{\Black{$-i\mu^{\epsilon/2}\left(Y_{e}\right)_{ij}\delta_{\alpha\beta}P_R$}}}
    \SetWidth{0.7}
    \SetColor{Black}
    \Line[arrow,arrowpos=0.5,arrowlength=1.5,arrowwidth=0.6,arrowinset=0.2,flip](386.718,-203.074)(428.085,-161.707)
    \Line[arrow,arrowpos=0.5,arrowlength=1.5,arrowwidth=0.6,arrowinset=0.2,flip](427.458,-161.707)(386.091,-120.34)
    \Line[dash,dashsize=4.387,arrow,arrowpos=0.5,arrowlength=1.5,arrowwidth=0.6,arrowinset=0.2](493.896,-161.707)(428.085,-161.707)
    \SetWidth{0.6}
    \SetColor{Gray}
    \Arc[arrow,arrowpos=0.5,arrowlength=1.5,arrowwidth=0.6,arrowinset=0.2](401.179,-161.757)(11.237,-70.691,63.755)
    \Text(371.049,-211.222)[lb]{\small{\Black{$l_{\alpha i}$}}}
    \Text(373.556,-112.192)[lb]{\small{\Black{$e_j$}}}
    \Text(475.093,-152.932)[lb]{\small{\Black{$\phi_{\beta}$}}}
  \end{picture}
}
\end{center}
{\bf dimension-5 effective vertex, $c^{(5)}$:}
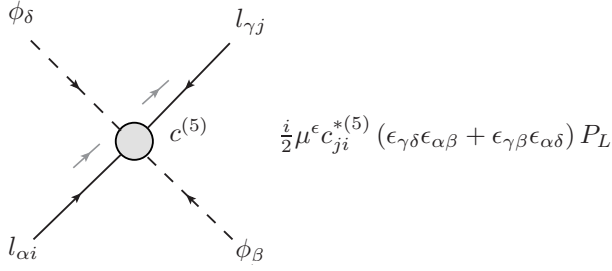
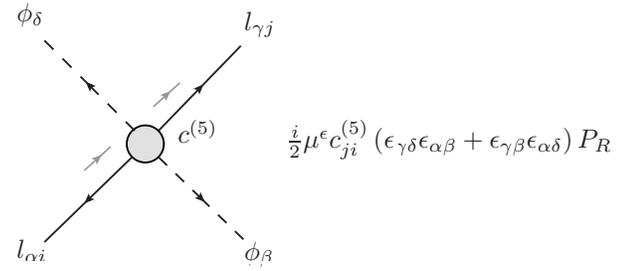
\begin{center}
\fcolorbox{white}{white}{
  \begin{picture}(391,108) (139,-140)
    \SetWidth{0.5}
    \SetColor{Black}
    \Text(495,-104)[lb]{\small{\Black{$\frac{i}{2}\mu^{\epsilon}c^{(5)}_{ji}\left(\epsilon_{\gamma\delta}\epsilon_{\alpha\beta} + \epsilon_{\gamma\beta}\epsilon_{\alpha\delta}\right)P_R$}}}
    \SetWidth{0.7}
    \GOval(440,-98)(7,7)(0){0.882}
    \Line[arrow,arrowpos=0.5,arrowlength=1.5,arrowwidth=0.6,arrowinset=0.2,flip](476,-61)(445,-93)
    \Line[arrow,arrowpos=0.5,arrowlength=1.5,arrowwidth=0.6,arrowinset=0.2,flip](402,-135)(435,-103)
    \Line[dash,dashsize=5,arrow,arrowpos=0.5,arrowlength=1.5,arrowwidth=0.6,arrowinset=0.2,flip](402,-59)(436,-93)
    \Line[dash,dashsize=5,arrow,arrowpos=0.5,arrowlength=1.5,arrowwidth=0.6,arrowinset=0.2,flip](477,-134)(445,-103)
    \Text(393,-143)[lb]{\small{\Black{$l_{\alpha i}$}}}
    \Text(479,-145)[lb]{\small{\Black{$\phi_{\beta}$}}}
    \Text(393,-54)[lb]{\small{\Black{$\phi_{\delta}$}}}
    \Text(479,-58)[lb]{\small{\Black{$l_{\gamma j}$}}}
    \SetWidth{0.6}
    \SetColor{Gray}
    \Line[arrow,arrowpos=0.5,arrowlength=1.5,arrowwidth=0.6,arrowinset=0.2](443,-84)(453,-75)
    \Line[arrow,arrowpos=0.5,arrowlength=1.5,arrowwidth=0.6,arrowinset=0.2](417,-108)(427,-99)
    \Text(454,-98)[lb]{\small{\Black{$c^{(5)}$}}}
    \Text(238,-102)[lb]{\small{\Black{$\frac{i}{2}\mu^{\epsilon}c^{\ast (5)}_{ji}\left(\epsilon_{\gamma\delta}\epsilon_{\alpha\beta} + \epsilon_{\gamma\beta}\epsilon_{\alpha\delta}\right)P_L$}}}
    \SetWidth{0.7}
    \SetColor{Black}
    \GOval(183,-97)(7,7)(0){0.882}
    \Line[arrow,arrowpos=0.5,arrowlength=1.5,arrowwidth=0.6,arrowinset=0.2](219,-60)(188,-92)
    \Line[arrow,arrowpos=0.5,arrowlength=1.5,arrowwidth=0.6,arrowinset=0.2](145,-134)(178,-102)
    \Line[dash,dashsize=5,arrow,arrowpos=0.5,arrowlength=1.5,arrowwidth=0.6,arrowinset=0.2](144,-59)(178,-93)
    \Line[dash,dashsize=5,arrow,arrowpos=0.5,arrowlength=1.5,arrowwidth=0.6,arrowinset=0.2](220,-134)(188,-103)
    \Text(136,-142)[lb]{\small{\Black{$l_{\alpha i}$}}}
    \Text(222,-144)[lb]{\small{\Black{$\phi_{\beta}$}}}
    \Text(136,-53)[lb]{\small{\Black{$\phi_{\delta}$}}}
    \Text(222,-57)[lb]{\small{\Black{$l_{\gamma j}$}}}
    \SetWidth{0.6}
    \SetColor{Gray}
    \Line[arrow,arrowpos=0.5,arrowlength=1.5,arrowwidth=0.6,arrowinset=0.2](186,-83)(196,-74)
    \Line[arrow,arrowpos=0.5,arrowlength=1.5,arrowwidth=0.6,arrowinset=0.2](160,-107)(170,-98)
    \Text(197,-97)[lb]{\small{\Black{$c^{(5)}$}}}
  \end{picture}
}
\end{center}
{\bf dimension-6 effective vertex, $c^{(6)}$:}
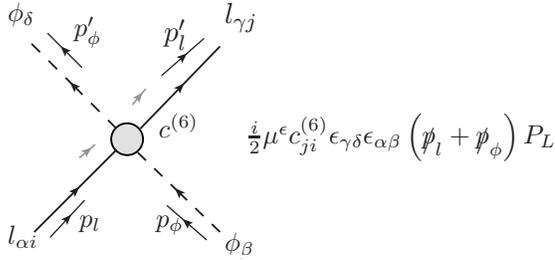
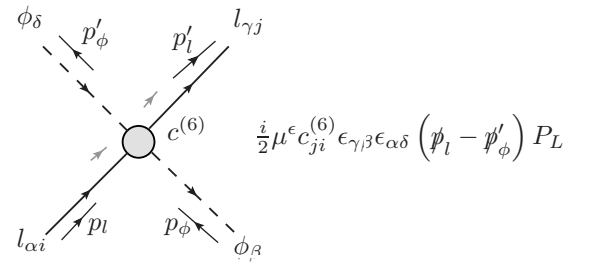
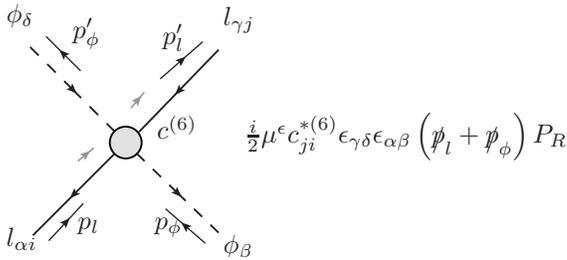
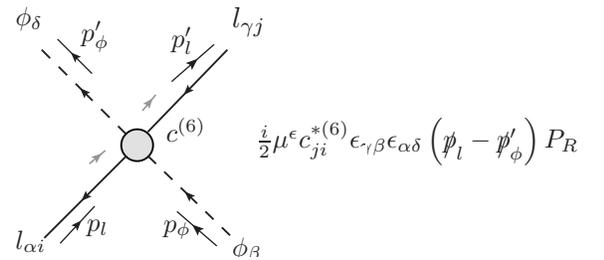
\begin{center}
\fcolorbox{white}{white}{
  \begin{picture}(395,105) (237,-257)
    \SetWidth{0.5}
    \SetColor{Black}
    \Text(324,-224)[lb]{\small{\Black{$\frac{i}{2}\mu^{\epsilon}c^{(6)}_{ji}\epsilon_{\gamma\delta}\epsilon_{\alpha\beta}\left(\slashed{p}_l+\slashed{p}_{\phi} \right)P_L$}}}
    \SetWidth{0.7}
    \GOval(278,-215)(6,6)(0){0.882}
    \Line[arrow,arrowpos=0.5,arrowlength=1.5,arrowwidth=0.6,arrowinset=0.2,flip](313,-180)(283,-211)
    \Line[arrow,arrowpos=0.5,arrowlength=1.5,arrowwidth=0.6,arrowinset=0.2](243,-250)(274,-219)
    \Line[dash,dashsize=5,arrow,arrowpos=0.5,arrowlength=1.5,arrowwidth=0.6,arrowinset=0.2,flip](242,-179)(274,-211)
    \Line[dash,dashsize=5,arrow,arrowpos=0.5,arrowlength=1.5,arrowwidth=0.6,arrowinset=0.2](313,-250)(282,-220)
    \Text(234,-257)[lb]{\small{\Black{$l_{\alpha i}$}}}
    \Text(316,-261)[lb]{\small{\Black{$\phi_{\beta}$}}}
    \Text(234,-173)[lb]{\small{\Black{$\phi_{\delta}$}}}
    \Text(316,-175)[lb]{\normalsize{\Black{$l_{\gamma j}$}}}
    \SetWidth{0.5}
    \Line[arrow,arrowpos=0.5,arrowlength=1.5,arrowwidth=0.6,arrowinset=0.2](291,-191)(307,-177)
    \Line[arrow,arrowpos=0.5,arrowlength=1.5,arrowwidth=0.6,arrowinset=0.2](261,-188)(248,-175)
    \Line[arrow,arrowpos=0.5,arrowlength=1.5,arrowwidth=0.6,arrowinset=0.2](249,-252)(262,-238)
    \Line[arrow,arrowpos=0.5,arrowlength=1.5,arrowwidth=0.6,arrowinset=0.2](308,-253)(293,-240)
    \Text(261,-251)[lb]{\small{\Black{$p_l$}}}
    \Text(290,-252)[lb]{\small{\Black{$p_{\phi}$}}}
    \Text(259,-181)[lb]{\small{\Black{$p_{\phi}^{\prime}$}}}
    \Text(293,-182)[lb]{\normalsize{\Black{$p_l^{\prime}$}}}
    \Text(291,-215)[lb]{\small{\Black{$c^{(6)}$}}}
    \SetWidth{0.6}
    \SetColor{Gray}
    \Line[arrow,arrowpos=0.5,arrowlength=1.5,arrowwidth=0.6,arrowinset=0.2](260,-222)(266,-217)
    \Line[arrow,arrowpos=0.5,arrowlength=1.5,arrowwidth=0.6,arrowinset=0.2](280,-202)(285,-196)
    \Text(597,-224)[lb]{\small{\Black{$\frac{i}{2}\mu^{\epsilon}c^{(6)}_{ji}\epsilon_{\gamma\beta}\epsilon_{\alpha\delta}\left(\slashed{p}_l-\slashed{p}_{\phi}^{\prime} \right)P_L$}}}
    \SetWidth{0.7}
    \SetColor{Black}
    \GOval(551,-216)(6,6)(0){0.882}
    \Line[arrow,arrowpos=0.5,arrowlength=1.5,arrowwidth=0.6,arrowinset=0.2,flip](585,-180)(555,-211)
    \Line[arrow,arrowpos=0.5,arrowlength=1.5,arrowwidth=0.6,arrowinset=0.2](516,-251)(547,-220)
    \Line[dash,dashsize=5,arrow,arrowpos=0.5,arrowlength=1.5,arrowwidth=0.6,arrowinset=0.2](515,-180)(547,-212)
    \Line[dash,dashsize=5,arrow,arrowpos=0.5,arrowlength=1.5,arrowwidth=0.6,arrowinset=0.2,flip](587,-250)(556,-220)
    \Text(507,-258)[lb]{\small{\Black{$l_{\alpha i}$}}}
    \Text(589,-262)[lb]{\small{\Black{$\phi_{\beta}$}}}
    \Text(507,-174)[lb]{\small{\Black{$\phi_{\delta}$}}}
    \Text(589,-176)[lb]{\small{\Black{$l_{\gamma j}$}}}
    \SetWidth{0.5}
    \Line[arrow,arrowpos=0.5,arrowlength=2,arrowwidth=0.8,arrowinset=0.2](564,-192)(580,-178)
    \Line[arrow,arrowpos=0.5,arrowlength=1.5,arrowwidth=0.6,arrowinset=0.2](534,-189)(521,-176)
    \Line[arrow,arrowpos=0.5,arrowlength=1.5,arrowwidth=0.6,arrowinset=0.2](522,-253)(535,-239)
    \Line[arrow,arrowpos=0.5,arrowlength=1.5,arrowwidth=0.6,arrowinset=0.2](581,-254)(566,-241)
    \Text(534,-252)[lb]{\small{\Black{$p_l$}}}
    \Text(563,-253)[lb]{\small{\Black{$p_{\phi}$}}}
    \Text(532,-182)[lb]{\small{\Black{$p_{\phi}^{\prime}$}}}
    \Text(566,-183)[lb]{\small{\Black{$p_l^{\prime}$}}}
    \Text(564,-216)[lb]{\small{\Black{$c^{(6)}$}}}
    \SetWidth{0.6}
    \SetColor{Gray}
    \Line[arrow,arrowpos=0.5,arrowlength=1.5,arrowwidth=0.6,arrowinset=0.2](533,-223)(539,-218)
    \Line[arrow,arrowpos=0.5,arrowlength=1.5,arrowwidth=0.6,arrowinset=0.2](553,-203)(558,-197)
  \end{picture}
}
\end{center}
\begin{center}
\fcolorbox{white}{white}{
  \begin{picture}(396,105) (236,-402)
    \SetWidth{0.5}
    \SetColor{Black}
    \Text(323,-368)[lb]{\small{\Black{$\frac{i}{2}\mu^{\epsilon}c^{\ast (6)}_{ji}\epsilon_{\gamma\delta}\epsilon_{\alpha\beta}\left(\slashed{p}_l+\slashed{p}_{\phi} \right)P_R$}}}
    \SetWidth{0.7}
    \GOval(277,-360)(6,6)(0){0.882}
    \Line[arrow,arrowpos=0.5,arrowlength=1.5,arrowwidth=0.6,arrowinset=0.2](312,-325)(282,-356)
    \Line[arrow,arrowpos=0.5,arrowlength=1.5,arrowwidth=0.6,arrowinset=0.2,flip](242,-395)(273,-364)
    \Line[dash,dashsize=5,arrow,arrowpos=0.5,arrowlength=1.5,arrowwidth=0.6,arrowinset=0.2](241,-324)(273,-356)
    \Line[dash,dashsize=5,arrow,arrowpos=0.5,arrowlength=1.5,arrowwidth=0.6,arrowinset=0.2,flip](312,-394)(281,-364)
    \Text(233,-402)[lb]{\small{\Black{$l_{\alpha i}$}}}
    \Text(315,-406)[lb]{\small{\Black{$\phi_{\beta}$}}}
    \Text(233,-318)[lb]{\small{\Black{$\phi_{\delta}$}}}
    \Text(315,-320)[lb]{\small{\Black{$l_{\gamma j}$}}}
    \SetWidth{0.5}
    \Line[arrow,arrowpos=0.5,arrowlength=1.5,arrowwidth=0.6,arrowinset=0.2](290,-336)(306,-322)
    \Line[arrow,arrowpos=0.5,arrowlength=1.5,arrowwidth=0.6,arrowinset=0.2](260,-333)(247,-320)
    \Line[arrow,arrowpos=0.5,arrowlength=1.5,arrowwidth=0.6,arrowinset=0.2](248,-397)(261,-383)
    \Line[arrow,arrowpos=0.5,arrowlength=1.5,arrowwidth=0.6,arrowinset=0.2](307,-398)(292,-385)
    \Text(260,-396)[lb]{\small{\Black{$p_l$}}}
    \Text(289,-397)[lb]{\small{\Black{$p_{\phi}$}}}
    \Text(258,-326)[lb]{\small{\Black{$p_{\phi}^{\prime}$}}}
    \Text(292,-327)[lb]{\small{\Black{$p_l^{\prime}$}}}
    \Text(290,-360)[lb]{\small{\Black{$c^{(6)}$}}}
    \SetWidth{0.6}
    \SetColor{Gray}
    \Line[arrow,arrowpos=0.5,arrowlength=1.5,arrowwidth=0.6,arrowinset=0.2](259,-367)(265,-362)
    \Line[arrow,arrowpos=0.5,arrowlength=1.5,arrowwidth=0.6,arrowinset=0.2](279,-347)(284,-341)
    \Text(597,-370)[lb]{\small{\Black{$\frac{i}{2}\mu^{\epsilon}c^{\ast (6)}_{ji}\epsilon_{\gamma\beta}\epsilon_{\alpha\delta}\left(\slashed{p}_l-\slashed{p}_{\phi}^{\prime} \right)P_R$}}}
    \SetWidth{0.7}
    \SetColor{Black}
    \GOval(550,-361)(6,6)(0){0.882}
    \Line[arrow,arrowpos=0.5,arrowlength=1.5,arrowwidth=0.6,arrowinset=0.2](584,-325)(554,-356)
    \Line[arrow,arrowpos=0.5,arrowlength=1.5,arrowwidth=0.6,arrowinset=0.2,flip](515,-396)(546,-365)
    \Line[dash,dashsize=5,arrow,arrowpos=0.5,arrowlength=1.5,arrowwidth=0.6,arrowinset=0.2,flip](514,-325)(546,-357)
    \Line[dash,dashsize=5,arrow,arrowpos=0.5,arrowlength=1.5,arrowwidth=0.6,arrowinset=0.2](585,-395)(554,-365)
    \Text(506,-403)[lb]{\small{\Black{$l_{\alpha i}$}}}
    \Text(588,-407)[lb]{\small{\Black{$\phi_{\beta}$}}}
    \Text(506,-319)[lb]{\small{\Black{$\phi_{\delta}$}}}
    \Text(588,-321)[lb]{\normalsize{\Black{$l_{\gamma j}$}}}
    \SetWidth{0.5}
    \Line[arrow,arrowpos=0.5,arrowlength=2,arrowwidth=0.8,arrowinset=0.2](563,-337)(579,-323)
    \Line[arrow,arrowpos=0.5,arrowlength=1.5,arrowwidth=0.6,arrowinset=0.2](533,-334)(520,-321)
    \Line[arrow,arrowpos=0.5,arrowlength=1.5,arrowwidth=0.6,arrowinset=0.2](521,-398)(534,-384)
    \Line[arrow,arrowpos=0.5,arrowlength=2,arrowwidth=0.8,arrowinset=0.2](580,-399)(565,-386)
    \Text(533,-397)[lb]{\small{\Black{$p_l$}}}
    \Text(562,-398)[lb]{\small{\Black{$p_{\phi}$}}}
    \Text(531,-327)[lb]{\small{\Black{$p_{\phi}^{\prime}$}}}
    \Text(565,-328)[lb]{\small{\Black{$p_l^{\prime}$}}}
    \Text(563,-361)[lb]{\small{\Black{$c^{(6)}$}}}
    \SetWidth{0.6}
    \SetColor{Gray}
    \Line[arrow,arrowpos=0.5,arrowlength=1.5,arrowwidth=0.6,arrowinset=0.2](532,-368)(538,-363)
    \Line[arrow,arrowpos=0.5,arrowlength=1.5,arrowwidth=0.6,arrowinset=0.2](552,-348)(557,-342)
  \end{picture}
}
\end{center}
\newpage
{\bf counterterm of $l_L$ and $\phi$:}
\begin{center}
\fcolorbox{white}{white}{
  \begin{picture}(451,50) (78,-152)
    \SetWidth{0.7}
    \SetColor{Black}
    \Line[arrow,arrowpos=0.5,arrowlength=1.5,arrowwidth=0.6,arrowinset=0.2](79.52,-130.214)(128.226,-130.214)
    \Text(79.52,-122.262)[lb]{\small{\Black{$l_{\alpha i}$}}}
    \Text(88.466,-154.07)[lb]{\small{\Black{$p$}}}
    \SetWidth{0.5}
    \Line[arrow,arrowpos=0.5,arrowlength=1.5,arrowwidth=0.6,arrowinset=0.2](80.514,-136.178)(104.37,-136.178)
    \SetWidth{0.7}
    \GOval(134.19,-130.214)(5.964,5.964)(0){0.882}
    \Line(131.208,-126.238)(137.172,-135.184)
    \Line(130.214,-134.19)(138.166,-126.238)
    \SetWidth{0.6}
    \SetColor{Gray}
    \Line[arrow,arrowpos=0.5,arrowlength=1.5,arrowwidth=0.6,arrowinset=0.2](124.25,-115.304)(144.13,-115.304)
    \Text(207.746,-133.196)[lb]{\small{\Black{$i\slashed{p}\left(\Delta Z_{l_L}\right)_{ij}P_L\delta_{\beta\alpha}$}}}
    \SetWidth{0.7}
    \SetColor{Black}
    \Line[arrow,arrowpos=0.5,arrowlength=1.5,arrowwidth=0.6,arrowinset=0.2](140.154,-130.214)(187.866,-130.214)
    \Text(171.962,-123.256)[lb]{\small{\Black{$l_{\beta j}$}}}
    \Text(172.956,-157.052)[lb]{\small{\Black{$p$}}}
    \SetWidth{0.5}
    \Line[arrow,arrowpos=0.5,arrowlength=1.5,arrowwidth=0.6,arrowinset=0.2](164.01,-137.172)(187.866,-137.172)
    \SetWidth{0.7}
    \Line[dash,dashsize=6.958,arrow,arrowpos=0.5,arrowlength=1.5,arrowwidth=0.6,arrowinset=0.2](370.762,-133.196)(418.474,-133.196)
    \Line[dash,dashsize=6.958,arrow,arrowpos=0.5,arrowlength=1.5,arrowwidth=0.6,arrowinset=0.2](431.396,-133.196)(479.108,-133.196)
    \Text(365.792,-125.244)[lb]{\small{\Black{$\phi_{\alpha}$}}}
    \Text(473.144,-125.244)[lb]{\small{\Black{$\phi_{\beta}$}}}
    \GOval(425.432,-133.196)(5.964,5.964)(0){0.882}
    \Line(422.45,-129.22)(428.414,-138.166)
    \Line(421.456,-137.172)(429.408,-129.22)
    \Text(495.012,-138.166)[lb]{\small{\Black{$i\left(p^2\Delta Z_{\phi}-\Delta m^2\right)\delta_{\beta\alpha}$}}}
  \end{picture}
}
\end{center}
{\bf $c^{(6)}$ counterterm:}
\begin{center}
\fcolorbox{white}{white}{
  \begin{picture}(394,105) (237,-257)
    \SetWidth{0.7}
    \SetColor{Black}
    \GOval(278,-215)(6,6)(0){0.882}
    \Text(316,-175)[lb]{\normalsize{\Black{$l_{\gamma j}$}}}
    \GOval(551,-216)(6,6)(0){0.882}
    \Line(273,-211)(283,-220)
    \Line(273,-220)(283,-211)
    \Line(546,-211)(556,-220)
    \Line(546,-220)(556,-211)
    \Text(325,-224)[lb]{\small{\Black{$\frac{i}{2}\mu^{\epsilon}\Delta c^{(6)}_{ji}\epsilon_{\gamma\delta}\epsilon_{\alpha\beta}\left(\slashed{p}_l+\slashed{p}_{\phi} \right)P_L$}}}
    \Line[arrow,arrowpos=0.5,arrowlength=1.5,arrowwidth=0.6,arrowinset=0.2,flip](313,-180)(283,-211)
    \Line[arrow,arrowpos=0.5,arrowlength=1.5,arrowwidth=0.6,arrowinset=0.2](243,-250)(274,-219)
    \Line[dash,dashsize=5,arrow,arrowpos=0.5,arrowlength=1.5,arrowwidth=0.6,arrowinset=0.2,flip](242,-179)(274,-211)
    \Line[dash,dashsize=5,arrow,arrowpos=0.5,arrowlength=1.5,arrowwidth=0.6,arrowinset=0.2](313,-250)(282,-220)
    \Text(234,-257)[lb]{\small{\Black{$l_{\alpha i}$}}}
    \Text(316,-261)[lb]{\small{\Black{$\phi_{\beta}$}}}
    \Text(234,-173)[lb]{\small{\Black{$\phi_{\delta}$}}}
    \SetWidth{0.5}
    \Line[arrow,arrowpos=0.5,arrowlength=1.5,arrowwidth=0.6,arrowinset=0.2](291,-191)(307,-177)
    \Line[arrow,arrowpos=0.5,arrowlength=1.5,arrowwidth=0.6,arrowinset=0.2](261,-188)(248,-175)
    \Line[arrow,arrowpos=0.5,arrowlength=1.5,arrowwidth=0.6,arrowinset=0.2](249,-252)(262,-238)
    \Line[arrow,arrowpos=0.5,arrowlength=1.5,arrowwidth=0.6,arrowinset=0.2](308,-253)(293,-240)
    \Text(261,-251)[lb]{\small{\Black{$p_l$}}}
    \Text(290,-252)[lb]{\small{\Black{$p_{\phi}$}}}
    \Text(259,-181)[lb]{\small{\Black{$p_{\phi}^{\prime}$}}}
    \Text(293,-182)[lb]{\normalsize{\Black{$p_l^{\prime}$}}}
    \Text(291,-215)[lb]{\small{\Black{$c^{(6)}$}}}
    \SetWidth{0.6}
    \SetColor{Gray}
    \Line[arrow,arrowpos=0.5,arrowlength=1.5,arrowwidth=0.6,arrowinset=0.2](260,-222)(266,-217)
    \Line[arrow,arrowpos=0.5,arrowlength=1.5,arrowwidth=0.6,arrowinset=0.2](280,-202)(285,-196)
    \Text(596,-223)[lb]{\small{\Black{$\frac{i}{2}\mu^{\epsilon}\Delta c^{(6)}_{ji}\epsilon_{\gamma\beta}\epsilon_{\alpha\delta}\left(\slashed{p}_l-\slashed{p}_{\phi}^{\prime} \right)P_L$}}}
    \SetWidth{0.7}
    \SetColor{Black}
    \Line[arrow,arrowpos=0.5,arrowlength=1.5,arrowwidth=0.6,arrowinset=0.2,flip](585,-180)(555,-211)
    \Line[arrow,arrowpos=0.5,arrowlength=1.5,arrowwidth=0.6,arrowinset=0.2](516,-251)(547,-220)
    \Line[dash,dashsize=5,arrow,arrowpos=0.5,arrowlength=1.5,arrowwidth=0.6,arrowinset=0.2](515,-180)(547,-212)
    \Line[dash,dashsize=5,arrow,arrowpos=0.5,arrowlength=1.5,arrowwidth=0.6,arrowinset=0.2,flip](587,-250)(556,-220)
    \Text(507,-258)[lb]{\small{\Black{$l_{\alpha i}$}}}
    \Text(589,-262)[lb]{\small{\Black{$\phi_{\beta}$}}}
    \Text(507,-174)[lb]{\small{\Black{$\phi_{\delta}$}}}
    \Text(589,-176)[lb]{\small{\Black{$l_{\gamma j}$}}}
    \SetWidth{0.5}
    \Line[arrow,arrowpos=0.5,arrowlength=2,arrowwidth=0.8,arrowinset=0.2](564,-192)(580,-178)
    \Line[arrow,arrowpos=0.5,arrowlength=1.5,arrowwidth=0.6,arrowinset=0.2](534,-189)(521,-176)
    \Line[arrow,arrowpos=0.5,arrowlength=1.5,arrowwidth=0.6,arrowinset=0.2](522,-253)(535,-239)
    \Line[arrow,arrowpos=0.5,arrowlength=1.5,arrowwidth=0.6,arrowinset=0.2](581,-254)(566,-241)
    \Text(534,-252)[lb]{\small{\Black{$p_l$}}}
    \Text(563,-253)[lb]{\small{\Black{$p_{\phi}$}}}
    \Text(532,-182)[lb]{\small{\Black{$p_{\phi}^{\prime}$}}}
    \Text(566,-183)[lb]{\small{\Black{$p_l^{\prime}$}}}
    \Text(564,-216)[lb]{\small{\Black{$c^{(6)}$}}}
    \SetWidth{0.6}
    \SetColor{Gray}
    \Line[arrow,arrowpos=0.5,arrowlength=1.5,arrowwidth=0.6,arrowinset=0.2](533,-223)(539,-218)
    \Line[arrow,arrowpos=0.5,arrowlength=1.5,arrowwidth=0.6,arrowinset=0.2](553,-203)(558,-197)
  \end{picture}
}
\end{center}
\begin{center}
\fcolorbox{white}{white}{
  \begin{picture}(396,105) (236,-402)
    \SetWidth{0.7}
    \SetColor{Black}
    \Line(272,-356)(282,-365)
    \Line(272,-365)(282,-356)
    \Line(545,-356)(555,-365)
    \Line(545,-365)(555,-356)
    \Text(322,-371)[lb]{\small{\Black{$\frac{i}{2}\mu^{\epsilon}\Delta c^{\ast (6)}_{ji}\epsilon_{\gamma\delta}\epsilon_{\alpha\beta}\left(\slashed{p}_l+\slashed{p}_{\phi} \right)P_R$}}}
    \GOval(277,-360)(6,6)(0){0.882}
    \Line[arrow,arrowpos=0.5,arrowlength=1.5,arrowwidth=0.6,arrowinset=0.2](312,-325)(282,-356)
    \Line[arrow,arrowpos=0.5,arrowlength=1.5,arrowwidth=0.6,arrowinset=0.2,flip](242,-395)(273,-364)
    \Line[dash,dashsize=5,arrow,arrowpos=0.5,arrowlength=1.5,arrowwidth=0.6,arrowinset=0.2](241,-324)(273,-356)
    \Line[dash,dashsize=5,arrow,arrowpos=0.5,arrowlength=1.5,arrowwidth=0.6,arrowinset=0.2,flip](312,-394)(281,-364)
    \Text(233,-402)[lb]{\small{\Black{$l_{\alpha i}$}}}
    \Text(315,-406)[lb]{\small{\Black{$\phi_{\beta}$}}}
    \Text(233,-318)[lb]{\small{\Black{$\phi_{\delta}$}}}
    \Text(315,-320)[lb]{\small{\Black{$l_{\gamma j}$}}}
    \SetWidth{0.5}
    \Line[arrow,arrowpos=0.5,arrowlength=1.5,arrowwidth=0.6,arrowinset=0.2](290,-336)(306,-322)
    \Line[arrow,arrowpos=0.5,arrowlength=1.5,arrowwidth=0.6,arrowinset=0.2](260,-333)(247,-320)
    \Line[arrow,arrowpos=0.5,arrowlength=1.5,arrowwidth=0.6,arrowinset=0.2](248,-397)(261,-383)
    \Line[arrow,arrowpos=0.5,arrowlength=1.5,arrowwidth=0.6,arrowinset=0.2](307,-398)(292,-385)
    \Text(260,-396)[lb]{\small{\Black{$p_l$}}}
    \Text(289,-397)[lb]{\small{\Black{$p_{\phi}$}}}
    \Text(258,-326)[lb]{\small{\Black{$p_{\phi}^{\prime}$}}}
    \Text(292,-327)[lb]{\small{\Black{$p_l^{\prime}$}}}
    \Text(290,-360)[lb]{\small{\Black{$c^{(6)}$}}}
    \SetWidth{0.6}
    \SetColor{Gray}
    \Line[arrow,arrowpos=0.5,arrowlength=1.5,arrowwidth=0.6,arrowinset=0.2](259,-367)(265,-362)
    \Line[arrow,arrowpos=0.5,arrowlength=1.5,arrowwidth=0.6,arrowinset=0.2](279,-347)(284,-341)
    \Text(597,-371)[lb]{\small{\Black{$\frac{i}{2}\mu^{\epsilon}\Delta c^{\ast (6)}_{ji}\epsilon_{\gamma\beta}\epsilon_{\alpha\delta}\left(\slashed{p}_l-\slashed{p}_{\phi}^{\prime} \right)P_R$}}}
    \SetWidth{0.7}
    \SetColor{Black}
    \GOval(550,-361)(6,6)(0){0.882}
    \Line[arrow,arrowpos=0.5,arrowlength=1.5,arrowwidth=0.6,arrowinset=0.2](584,-325)(554,-356)
    \Line[arrow,arrowpos=0.5,arrowlength=1.5,arrowwidth=0.6,arrowinset=0.2,flip](515,-396)(546,-365)
    \Line[dash,dashsize=5,arrow,arrowpos=0.5,arrowlength=1.5,arrowwidth=0.6,arrowinset=0.2,flip](514,-325)(546,-357)
    \Line[dash,dashsize=5,arrow,arrowpos=0.5,arrowlength=1.5,arrowwidth=0.6,arrowinset=0.2](585,-395)(554,-365)
    \Text(506,-403)[lb]{\small{\Black{$l_{\alpha i}$}}}
    \Text(588,-407)[lb]{\small{\Black{$\phi_{\beta}$}}}
    \Text(506,-319)[lb]{\small{\Black{$\phi_{\delta}$}}}
    \Text(588,-321)[lb]{\normalsize{\Black{$l_{\gamma j}$}}}
    \SetWidth{0.5}
    \Line[arrow,arrowpos=0.5,arrowlength=2,arrowwidth=0.8,arrowinset=0.2](563,-337)(579,-323)
    \Line[arrow,arrowpos=0.5,arrowlength=1.5,arrowwidth=0.6,arrowinset=0.2](533,-334)(520,-321)
    \Line[arrow,arrowpos=0.5,arrowlength=1.5,arrowwidth=0.6,arrowinset=0.2](521,-398)(534,-384)
    \Line[arrow,arrowpos=0.5,arrowlength=2,arrowwidth=0.8,arrowinset=0.2](580,-399)(565,-386)
    \Text(533,-397)[lb]{\small{\Black{$p_l$}}}
    \Text(562,-398)[lb]{\small{\Black{$p_{\phi}$}}}
    \Text(531,-327)[lb]{\small{\Black{$p_{\phi}^{\prime}$}}}
    \Text(565,-328)[lb]{\small{\Black{$p_l^{\prime}$}}}
    \Text(563,-361)[lb]{\small{\Black{$c^{(6)}$}}}
    \SetWidth{0.6}
    \SetColor{Gray}
    \Line[arrow,arrowpos=0.5,arrowlength=1.5,arrowwidth=0.6,arrowinset=0.2](532,-368)(538,-363)
    \Line[arrow,arrowpos=0.5,arrowlength=1.5,arrowwidth=0.6,arrowinset=0.2](552,-348)(557,-342)
    \SetWidth{0.7}
    \SetColor{Black}
    \Line(272,-356)(282,-365)
    \Line(272,-365)(282,-356)
    \Line(545,-356)(554,-365)
    \Line(545,-366)(555,-355)
  \end{picture}
}
\end{center}
\section{Appendix B}
Here we give the divergent part of different diagrams\\

{\bf Self energy diagrams of $l_L$, Fig. \ref{lep-slf-enrg}:}\\
(I)
\begin{eqnarray}
\frac{i}{16\pi^2}\left(Y_{\nu} Y_{\nu}^{\dag}\right)_{ji}\delta_{\gamma\alpha}\slashed{p}P_L\frac{1}{\epsilon} + \text{UV finite} \nonumber
\end{eqnarray}
(II)
\begin{eqnarray}
\frac{i}{16\pi^2}\left(Y_e Y_e^{\dag}\right)_{ji}\delta_{\gamma\alpha}\slashed{p}P_L\frac{1}{\epsilon} + \text{UV finite} \nonumber
\end{eqnarray}
(III)
\begin{eqnarray}
\frac{i}{16\pi^2}\frac{g_1^2}{2}\delta_{ji}\delta_{\gamma\alpha}\xi_B\slashed{p}P_L\frac{1}{\epsilon} + \text{UV finite} \nonumber
\end{eqnarray}
(IV)
\begin{eqnarray}
\frac{i}{16\pi^2}\frac{3}{2}g_2^2\delta_{ji}\delta_{\gamma\alpha}\xi_W\slashed{p}P_L\frac{1}{\epsilon} + \text{UV finite} \nonumber
\end{eqnarray}
{\bf Self energy diagrams of $\phi$, Fig. \ref{higgs-slf-enrg}:}\\
(I)
\begin{eqnarray}
\frac{i}{16\pi^2}3\lambda m^2\delta_{\beta\alpha}\frac{1}{\epsilon} + \text{UV finite} \nonumber
\end{eqnarray}
(II)
\begin{eqnarray}
\frac{i}{16\pi^2}\frac{g_1^2}{2} \delta_{\beta\alpha}\left[\left(-3+\xi_B\right)p^2-\xi_B m^2\right]\frac{1}{\epsilon} + \text{UV finite} \nonumber
\end{eqnarray}
(III)
\begin{eqnarray}
\frac{i}{16\pi^2}\frac{3}{2}g_2^2 \delta_{\beta\alpha}\left[\left(-3+\xi_W\right)p^2-\xi_W m^2\right]\frac{1}{\epsilon} + \text{UV finite} \nonumber
\end{eqnarray}
(V)
\begin{eqnarray}
\frac{i}{16\pi^2}2\left(Y_{\nu}\right)_{jk} \left(Y_{\nu}^{\dag}\right)_{ki} \delta_{\beta\alpha}\left[p^2 - 2M_k^2\right]\frac{1}{\epsilon} + \text{UV finite} \nonumber
\end{eqnarray}
(VI)
\begin{eqnarray}
\frac{i}{16\pi^2}2\text{Tr}\left(Y_e Y_e^{\dag}\right)\delta_{\beta\alpha}p^2 \frac{1}{\epsilon} + \text{UV finite} \nonumber
\end{eqnarray}
(VII)
\begin{eqnarray}
\frac{i}{16\pi^2}6\text{Tr}\left(Y_u Y_u^{\dag}\right)\delta_{\beta\alpha}p^2 \frac{1}{\epsilon} + \text{UV finite} \nonumber
\end{eqnarray}
(VIII)
\begin{eqnarray}
\frac{i}{16\pi^2}6\text{Tr}\left(Y_d Y_d^{\dag}\right)\delta_{\beta\alpha}p^2 \frac{1}{\epsilon} + \text{UV finite} \nonumber
\end{eqnarray}
{\bf One loop correction of $c^{(6)}$, Fig. \ref{c6-vrtx}:}\\
(I)
\begin{eqnarray}
\frac{i}{16\pi^2}\mu^{\epsilon}\left(c^{(6)}Y_e Y_e^{\dag}\right)_{ji}\epsilon_{\gamma\beta}\epsilon_{\alpha\delta}\left(\slashed{p}_l-\slashed{p}_{\phi}^{\prime}\right)P_L\frac{1}{\epsilon} + \text{UV finite} \nonumber
\end{eqnarray}
(II)
\begin{eqnarray}
\frac{i}{16\pi^2}\mu^{\epsilon}\left(Y_e Y_e^{\dag}c^{(6)}\right)_{ji}\epsilon_{\gamma\beta}\epsilon_{\alpha\delta}\left(\slashed{p}_l-\slashed{p}_{\phi}^{\prime}\right)P_L\frac{1}{\epsilon} + \text{UV finite} \nonumber
\end{eqnarray}
(III)
\begin{eqnarray}
\frac{i}{16\pi^2}\mu^{\epsilon}\left(c^{(5)}c^{\ast(5)}\right)_{ji}\left(\epsilon_{\gamma\beta}\epsilon_{\alpha\delta} + \frac{1}{4}\delta_{\beta\delta}\delta_{\gamma\alpha}\right)\left(\slashed{p}_l-\slashed{p}_{\phi}^{\prime}\right)P_L\frac{1}{\epsilon} + \text{UV finite} \nonumber
\end{eqnarray}
(IV)
\begin{eqnarray}
\frac{i}{16\pi^2}\mu^{\epsilon}\frac{g_1^2}{4}c^{(6)}_{ji}\epsilon_{\gamma\beta}\epsilon_{\alpha\delta}\left(\slashed{p}_l-\slashed{p}_{\phi}^{\prime}\right)P_L\xi_B \frac{1}{\epsilon} + \text{UV finite} \nonumber
\end{eqnarray}
Corresponding $W$ boson diagram is three times the $B$ diagram with $g_1$ replaced by $g_2$.\\
(V)
\begin{eqnarray}
\frac{i}{16\pi^2}\mu^{\epsilon}\frac{g_1^2}{4}c^{(6)}_{ji}\epsilon_{\gamma\beta}\epsilon_{\alpha\delta}\left(\slashed{p}_l-\slashed{p}_{\phi}^{\prime}\right)P_L\xi_B \frac{1}{\epsilon} + \text{UV finite} \nonumber
\end{eqnarray}
Corresponding $W$ boson diagram is three times the $B$ diagram with $g_1$ replaced by $g_2$.\\
(VI)
\begin{eqnarray}
\frac{i}{16\pi^2}\mu^{\epsilon}\frac{g_1^2}{4}c^{(6)}_{ji}\epsilon_{\gamma\beta}\epsilon_{\alpha\delta}\left[\frac{3}{2}\slashed{p}_{\phi} + \xi_B\slashed{p}_{\phi}^{\prime}\right]P_L \frac{1}{\epsilon} + \text{UV finite} \nonumber
\end{eqnarray}
Corresponding $W$ boson diagram is identical with the $B$ diagram with $g_1$ replaced by $g_2$.\\
(VII)
\begin{eqnarray}
\frac{i}{16\pi^2}\mu^{\epsilon}\frac{g_1^2}{4}c^{(6)}_{ji}\epsilon_{\gamma\beta}\epsilon_{\alpha\delta}\left[\left(\frac{3}{2}+\xi_B\right)\slashed{p}_{\phi}^{\prime} + \xi_B\slashed{p}_l^{\prime}-\xi_B\slashed{p}_l\right]P_L \frac{1}{\epsilon} + \text{UV finite} \nonumber
\end{eqnarray}
Corresponding $W$ boson diagram is identical with the $B$ diagram with $g_1$ replaced by $g_2$.\\
(VIII)
\begin{eqnarray}
\frac{i}{16\pi^2}\mu^{\epsilon}\frac{g_1^2}{4}c^{(6)}_{ji}\epsilon_{\gamma\beta}\epsilon_{\alpha\delta}\xi_B \left(\slashed{p}_l-\slashed{p}_{\phi}^{\prime}\right)P_L\frac{1}{\epsilon} + \text{UV finite} \nonumber
\end{eqnarray}
Corresponding $W$ boson diagram is identical with the $B$ diagram with $g_1$ replaced by $g_2$.\\
(IX)
\begin{eqnarray}
\frac{i}{16\pi^2}\mu^{\epsilon}\left(-\frac{g_1^2}{4}\right)c^{(6)}_{ji}\epsilon_{\gamma\beta}\epsilon_{\alpha\delta}\xi_B\left(\slashed{p}_l^{\prime}+\slashed{p}_{\phi}^{\prime}\right)P_L\frac{1}{\epsilon} + \text{UV finite} \nonumber
\end{eqnarray}
Corresponding $W$ boson diagram is identical with the $B$ diagram with $g_1$ replaced by $g_2$.\\
(X)
\begin{eqnarray}
-\frac{i}{16\pi^2}\mu^{\epsilon}\frac{\lambda}{8}c^{(6)}_{ji}\left(\delta_{\gamma\alpha}\delta_{\beta\delta}+\epsilon_{\gamma\beta}\epsilon_{\alpha\delta}\right)\left(\slashed{p}_l + \slashed{p}_l^{\prime}\right)P_L\frac{1}{\epsilon} + \text{UV finite} \nonumber
\end{eqnarray}
In the above expressions, a term proportional to $\delta_{\gamma\alpha}\delta_{\beta\delta}$ arises in cases of diagram III and all the $W$ boson diagrams. We used the identity
\begin{eqnarray}
\epsilon_{\gamma\beta}\epsilon_{\alpha\delta} &=& \delta_{\gamma\alpha}\delta_{\beta\delta} - \delta_{\beta\alpha}\delta_{\gamma\delta}\nonumber \\
\Rightarrow \delta_{\gamma\alpha}\delta_{\beta\delta} &=& \epsilon_{\gamma\beta}\epsilon_{\alpha\delta} + \delta_{\beta\alpha}\delta_{\gamma\delta}\nonumber
\end{eqnarray}
to change $\delta_{\gamma\alpha}\delta_{\beta\delta}$ to $\epsilon_{\gamma\beta}\epsilon_{\alpha\delta}$. The other part ({\it i.e.} $\delta_{\beta\alpha}\delta_{\gamma\delta})$ is orthogonal to the non-unitary operator \cite{dimension6_1}.

\end{document}